\documentclass[a4paper,11pt]{article}

\usepackage{jheppub} 
\usepackage{graphicx}
\usepackage{color}
\usepackage{booktabs}
\usepackage{hyperref}
\usepackage{dsfont}
\usepackage{here}
\usepackage{subcaption}

\newcommand{\kslash}{k\kern-1ex /}
\newcommand{\pslash}{p\kern-1ex /}
\newcommand{\qslash}{q\kern-1ex /}
\newcommand{\lslash}{l\kern-1ex /}
\newcommand{\sslash}{s\kern-1ex /}
\newcommand{\Dslash}{D\kern-1.2ex /}

\newcommand{\beqa}{\begin{eqnarray}}
\newcommand{\eeqa}{\end{eqnarray}}
\newcommand{\Tr}{{\rm Tr}}

\newcommand{\be}{\begin{equation}}
\newcommand{\ee}{\end{equation}}
\newcommand{\bd}{\begin{description}}
\newcommand{\ed}{\end{description}}

\newcommand{\ben}{\begin{eqnarray}}
\newcommand{\een}{\end{eqnarray}}

\def\lsim{\raise0.3ex\hbox{$<$\kern-0.75em\raise-1.1ex\hbox{$\sim$}}}
\def\gsim{\raise0.3ex\hbox{$>$\kern-0.75em\raise-1.1ex\hbox{$\sim$}}}
\def\simgt{\rlap{\lower 3.5 pt\hbox{$\mathchar \sim$}}\raise 2.0pt \hbox {$>$}}
\def\simlt{\rlap{\lower 3.5 pt\hbox{$\mathchar \sim$}}\raise 2.0pt \hbox {$<$}}

\newcommand{\im}{\mathrm{i}}
\newcommand{\diff}{\mathrm{d}}

\newcommand{\e}{\mathrm{e}}

\begin{document}

\title{Tensor renormalization group study of (1+1)-dimensional U(1) gauge-Higgs model at $\theta=\pi$ with L{\"u}scher's admissibility condition}

\author[a,b]{Shinichiro Akiyama,}
    \affiliation[a]{Center for Computational Sciences, University of Tsukuba, Tsukuba, Ibaraki 305-8577, Japan}
    \affiliation[b]{Graduate School of Science, The University of Tokyo, Bunkyo-ku, Tokyo 113-0033, Japan}
    \emailAdd{akiyama@ccs.tsukuba.ac.jp}

\author[a]{Yoshinobu Kuramashi}
    \emailAdd{kuramasi@het.ph.tsukuba.ac.jp}

\abstract{
  We investigate the phase structure of the (1+1)-dimensional U(1) gauge-Higgs model with a $\theta$ term, where the U(1) gauge action is constructed with L{\"u}scher's admissibility condition. 
  Using the tensor renormalization group, both the complex action problem and topological freezing problem in the standard Monte Carlo simulation are avoided.
  We find the first-order phase transition with sufficiently large Higgs mass at $\theta=\pi$, where the $\mathds{Z}_2$ charge conjugation symmetry is spontaneously broken. 
  On the other hand, the symmetry is restored with a sufficiently small mass. 
  We determine the critical endpoint as a function of the Higgs mass parameter and show the critical behavior is in the two-dimensional Ising universality class.
}
\date{\today}

\preprint{UTHEP-789, UTCCS-P-157}

\maketitle

\section{Introduction}
\label{sec:intro}

At the end of the last century, L{\"u}scher introduced an admissibility condition for the gauge fields to be separated into disconnected subspaces corresponding to topological charges in the continuum theory~\cite{Luscher:1998du}:
\begin{align}
    \| 1-P_{\mu\nu}(n)\| < \epsilon \;\; {\rm for}\;\; ^\forall n,\mu,\nu,
\end{align}
where $\epsilon$ is a positive constant and $P_{\mu\nu}(n)$ is a product of link variables $U_{\mu}(n)$ as in the standard way,
\begin{align}
\label{eq:gauge_plaquette}
    P_{\mu\nu}(n)
    =
    U_{\mu}(n)
    U_{\nu}(n+{\hat \mu})
    U_{\mu}^\dagger(n+{\hat \nu}
    )U_{\nu}^\dagger(n)
    .
\end{align}
The link variable $U_{\mu}(n)$ lives on the link connecting the sites $n$ and $n+\hat{\mu}$.
As an example, he proposed the following gauge action to make the link variables satisfy the above condition:
\begin{align}
\label{eq:luscheraction}
    \beta S_{g} =
    \begin{cases}
        \displaystyle
        \beta\sum_{n,\mu>\nu}
        \frac{1-{\rm Re}P_{\mu\nu}(n)}{1-\| 1-P_{\mu\nu}(n)\|/\epsilon}\;\; 
        & {\rm if}\;\; \| 1-P_{\mu\nu}(n) \| < \epsilon,\\
        \infty & {\rm otherwise},
    \end{cases}
\end{align}
with the inverse gauge coupling $\beta$.
This action should have an advantage in investigating the topological effects of the gauge theories. 
However, early numerical studies with this action revealed that the topological change is substantially suppressed in the Monte Carlo histories~\cite{Fukaya:2003ph} and it is difficult to evaluate the contributions from different topological sectors.
As long as the Monte Carlo method is employed, the possible way is to perform calculations in the fixed topological sectors~\cite{Fukaya:2003ph,Fukaya:2004kp,Bietenholz:2005rd,Bietenholz:2016ymo} or to utilize open boundary conditions, dismissing the translational invariance of the system~\cite{Luscher:2011kk}.

However, this problem is potentially solved by the tensor renormalization group (TRG) method.~\footnote{
In this paper, the ``TRG method" or the ``TRG approach" refers to not only the original numerical algorithm proposed by Levin and Nave~\cite{Levin:2006jai} but also its extensions \cite{Gu:2010yh,PhysRevB.86.045139,Shimizu:2014uva,Sakai:2017jwp,Adachi:2019paf,Kadoh:2019kqk,Akiyama:2020soe}.
}
The major advantages of the TRG method over the Monte Carlo simulation are (i) no sign problem~\cite{Denbleyker:2013bea,Shimizu:2014fsa,Kawauchi:2016xng,Shimizu:2017onf,Unmuth-Yockey:2018ugm,Kadoh:2019ube,Kuramashi:2019cgs,Butt:2019uul,Bloch:2021uup,Takeda:2021mnc,Bloch:2021mjw,Nakayama:2021iyp,Hirasawa:2021qvh}, 
(ii) logarithmic computational cost on the system size, 
(iii) direct manipulation of the Grassmann variables~\cite{Gu:2010yh,Shimizu:2014uva,Takeda:2014vwa,Sakai:2017jwp,Yoshimura:2017jpk,Kadoh:2018hqq,Akiyama:2021xxr,Bloch:2022vqz,Akiyama:2023lvr,Yosprakob:2023tyr,Asaduzzaman:2023pyz}, 
and (iv) evaluation of the partition function or the path integral itself. 
The advantage (iv) ensures that the TRG calculation automatically includes full contributions from different topological sectors.
Moreover, the TRG method assumes the translational invariance of the system and can easily impose periodic boundary conditions.

\begin{figure}[htbp]
    \centering
    \includegraphics[width=0.7\hsize]{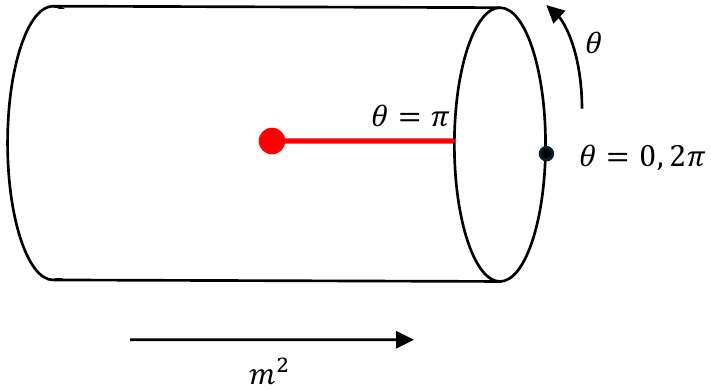}
    \caption{
    Schematic phase diagram of (1+1)$d$ U(1) gauge-Higgs model with a $\theta$ term. The horizontal axis denotes the Higgs mass-squared. 
    The red line denotes the first-order phase transition, which terminates at the critical endpoint expressed by the red blob.}
    \label{fig:phasedgm}
\end{figure}

In this paper, we investigate the phase structure of the (1+1)-dimensional ((1+1)$d$) U(1) gauge-Higgs model with a $\theta$ term, where the topological effects play an essential role, employing L{\"u}scher's gauge action of Eq.~\eqref{eq:luscheraction}. 
The Monte Carlo simulation of this model is extremely difficult due to a double whammy of the complex action problem and the topological freezing. 
Figure~\ref{fig:phasedgm} illustrates the expected phase diagram~\cite{Komargodski:2017dmc}. 
The model exhibits the first-order phase transition at $\theta=\pi$, where the $\mathds{Z}_2$ charge conjugation symmetry is spontaneously broken in the large positive Higgs mass-squared regime, including the pure gauge limit. 
\footnote{
There are several earlier studies on the $2d$ pure gauge theory with a $\theta$ term by the density of state approach~\cite{Gattringer:2020mbf}, complex Langevin method~\cite{Hirasawa:2020bnl}, and TRG~\cite{Kuramashi:2019cgs,Hirasawa:2021qvh}.
}
Once the Higgs mass-squared is sufficiently reduced, the symmetry is restored.  
We determine the critical endpoint as a function of the Higgs mass-squared and show its critical behavior is in the 2$d$ Ising universality class based on the numerical analysis of the transfer matrix and topological charge density.
We also compare our results with the previous work employing the dual lattice simulation based on the Villain gauge action, which is a non-compact gauge action on the lattice~\cite{Gattringer:2018dlw}.

This paper is organized as follows. 
In Sec.~\ref{sec:method}, we define the U(1) gauge-Higgs model with a $\theta$ term on a (1+1)$d$ lattice. 
We also demonstrate how to represent the path integral as a tensor network.
In Sec.~\ref{sec:results}, we first present the results for the pure U(1) gauge action which corresponds to the infinitely heavy limit of the Higgs mass, where the first-order phase transition takes place at $\theta=\pi$. 
After that, we discuss the phase transition with the finite lattice Higgs mass and determine the critical endpoint and its universality class. 
Section~\ref{sec:summary} is devoted to summary and outlook.

\section{Tensor network formulation}
\label{sec:method}

\subsection{The model on a lattice}

The U(1) gauge-Higgs model with a $\theta$ term is defined by
\begin{align}
\label{eq:action}
    S=
    \beta S_{g}+S_{h}+S_{\theta}
    .
\end{align}
We always consider the model on a square lattice with periodic boundary conditions.
We employ the L\"{u}scher gauge action,
\begin{align}
\label{eq:gauge_luscher}
    S_{g}
    =
    \begin{cases}
        \displaystyle
        \sum_{n}
        \frac{1-{\rm Re}P_{12}(n)}{1-[1-{\rm Re}P_{12}(n)]/\epsilon}& {\rm if~``admissible"} \\
        \infty & {\rm otherwise}
    \end{cases}
    ,
\end{align}
where $P_{12}(n)$ is defined by Eq.~\eqref{eq:gauge_plaquette} with $U_{\nu}(n)={\rm e}^{\im\vartheta_{\nu}(n)}$ and $\vartheta_{\nu}(n)\in[-\pi,\pi]$.
The admissibility condition is given by
\begin{align}
\label{eq:s_luscher}
    1-{\rm Re}P_{12}(n) < \epsilon.
\end{align}
When this condition is satisfied, the corresponding gauge fields are called admissible.
The admissibility condition makes gauge fields smooth and unphysical configurations are suppressed.
The space of admissible gauge fields is separated into disconnected subspaces which are labeled by the integers corresponding to topological charges in the continuum~\cite{Luscher:1998du}.
The Higgs part is defined by
\begin{align}
    S_{h}
    &=
    -
    \sum_{n}
    \sum_{\nu}
    \left[
        \phi^{*}(n)
        U_{\nu}(n)
        \phi(n+\hat{\nu})
        +
        \phi^{*}(n+\hat{\nu})
        U^{*}_{\nu}(n)
        \phi(n)
    \right]
    \nonumber\\
    &+M\sum_{n}\left|\phi(n)\right|^{2}
    +\lambda\sum_{n}\left|\phi(n)\right|^{4}
    .
\end{align}
The complex-valued Higgs fields are denoted by $\phi(n)$ and $M=m^{2}+4$ is the lattice Higgs mass where $m$ corresponds to the Higgs mass parameter in the continuum action.
The quartic coupling constant is denoted by $\lambda$.
Finally, the $\theta$ term is defined by
\begin{align}
\label{eq:theta_log}
    S_{\theta}
    =
    -\frac{\im\theta}{2\pi}
    \sum_{n}\ln P_{12}(n)
    .
\end{align}

\subsection{Tensor network formulation}

We consider the tensor network representation of the path integral defined as
\begin{align}
    Z=
    \prod_{n,\nu}\int^{\pi}_{-\pi}\frac{\diff\vartheta_{\nu}(n)}{2\pi}
    \prod_{n}\int_{\mathds{C}}\frac{\diff\phi(n)}{2\pi}
    \exp(-S)
    .
\end{align}
Parametrizing the complex-valued Higgs field by $\phi(n)=r(n)\e^{\im\varphi(n)}$, the integral measure is represented as
\begin{align}
    \int_{\mathds{C}}\frac{\diff\phi(n)}{2\pi}
    =
    \int^{\infty}_{0}r(n)\diff r(n)
    \int^{\pi}_{-\pi}\frac{\diff\varphi(n)}{2\pi}
    ,
\end{align}
and $S_{h}$ reads
\begin{align}
\label{eq:s_h_original}
    S_{h}
    &=
    -
    \sum_{n}
    \sum_{\nu}
    2r(n)r(n+\hat{\nu})
    \cos\left[\varphi(n+\hat{\nu})-\varphi(n)+\vartheta_{\nu}(n)\right]
    +\sum_{n}\left[Mr(n)^{2}+\lambda r(n)^{4}\right]
    .
\end{align}
We further introduce $\ell(n)=r(n)^{2}$ and rewrite Eq.~\eqref{eq:s_h_original} as
\begin{align}
    S_{h}
    &=
    -
    \sum_{n}
    \sum_{\nu}
    2\sqrt{\ell(n)\ell(n+\hat{\nu})}
    \cos\left[\varphi(n+\hat{\nu})-\varphi(n)+\vartheta_{\nu}(n)\right]
    +\sum_{n}\left[M\ell(n)+\lambda \ell(n)^{2}\right]
    .
\end{align}
Using the invariance of the Haar measure, or choosing the unitary gauge, we can eliminate $\varphi(n)$ from the path integral,
\begin{align}
    Z=
    \prod_{n,\nu}\int^{\pi}_{-\pi}\frac{\diff\vartheta_{\nu}(n)}{2\pi}
    \prod_{n}\int^{\infty}_{0}\frac{\diff\ell(n)}{2}
    \exp\left[-\beta S_{g}-S'_{h}-S_{\theta}\right]
    ,
\end{align}
with
\begin{align}
    S'_{h}
    &=
    -
    \sum_{n}
    \sum_{\nu}
    2\sqrt{\ell(n)\ell(n+\hat{\nu})}
    \cos\vartheta_{\nu}(n)
    +\sum_{n}\left[M\ell(n)+\lambda \ell(n)^{2}\right]
    .
\end{align}

In this study, we use the Gauss-Laguerre quadrature rule to discretize the integral over $\ell(n)$ and the Gauss-Legendre one for $\vartheta_{\nu}(n)$.
The efficacy of these Gauss quadrature rules has been reported in the previous TRG studies for the U(1) pure gauge theory with a $\theta$ term~\cite{Kuramashi:2019cgs} and complex $\phi^{4}$ theories~\cite{Kadoh:2019ube,Akiyama:2020ntf}.
\footnote{Another way to discretize the integral would be the character expansion. Although there is no systematic comparative study between the Gauss quadrature and the character expansion, a clear advantage of the former is that it can be easily applied to more complicated forms of lattice actions.}
The path integral is now approximated by
\begin{align}
\label{eq:z_approx}
    Z
    &\simeq
    Z(K_{g},K_{h})\nonumber\\
    &=
    \prod_{n,\nu}\sum_{\tilde{\vartheta}_{\nu}(n)\in D_{g}}
    \frac{w_{\tilde{\vartheta}_{\nu}(n)}}{2}
    \prod_{n}\sum_{\tilde{\ell}(n)\in D_{h}}
    \frac{w_{\tilde{\ell}(n)}\e^{\tilde{\ell}(n)}}{2}
    \exp\left[-\beta \tilde{S}_{g}-\tilde{S}'_{h}-\tilde{S}_{\theta}\right]
    ,
\end{align}
where
\begin{align}
    \tilde{S}_{g}
    =
    \begin{cases}
        \displaystyle
        \sum_{n}
        \frac{1-\cos\pi\left(\tilde{\vartheta}_{1}(n)+\tilde{\vartheta}_{2}(n+\hat{1})-\tilde{\vartheta}_{1}(n+\hat{2})-\tilde{\vartheta}_{2}(n)\right)}
        {1-\left[1-\cos\pi\left(\tilde{\vartheta}_{1}(n)+\tilde{\vartheta}_{2}(n+\hat{1})-\tilde{\vartheta}_{1}(n+\hat{2})-\tilde{\vartheta}_{2}(n)\right)\right]/\epsilon} & {\rm if~admissible} \\
        \infty & {\rm otherwise}
    \end{cases}
    ,
\end{align}
\begin{align}
    \tilde{S}'_{h}
    &=
    -
    \sum_{n}
    \sum_{\nu}
    2\sqrt{\tilde{\ell}(n)\tilde{\ell}(n+\hat{\nu})}
    \cos\pi\tilde{\vartheta}_{\nu}(n)
    +\sum_{n}\left[M\tilde{\ell}(n)+\lambda \tilde{\ell}(n)^{2}\right]
    ,
\end{align}
\begin{align}
    \tilde{S}_{\theta}
    =
    -\frac{\im\theta}{2\pi}
    \sum_{n}\ln\left[{\rm e}^{\im\pi\left(\tilde{\vartheta}_{1}(n)+\tilde{\vartheta}_{2}(n+\hat{1})-\tilde{\vartheta}_{1}(n+\hat{2})-\tilde{\vartheta}_{2}(n)\right)}\right]
    .
\end{align}
In Eq.~\eqref{eq:z_approx}, $\tilde{\ell}(n)$ denotes the sampling point according to the Gauss-Laguerre quadrature and $w_{\tilde{\ell}(n)}$ is the corresponding weight.
The number of sampling points in $D_{h}$ is denoted by $K_{h}$.
Similarly, $\tilde{\vartheta}_{\nu}(n)$ denotes the sampling point according to the Gauss-Legendre quadrature and $w_{\tilde{\vartheta}_{\nu}(n)}$ is the corresponding weight.
The number of sampling points in $D_{g}$ is denoted by $K_{g}$.
In the limits of $K_{g}\to\infty$ and $K_{h}\to\infty$, the original path integral is restored.
Eq.~\eqref{eq:z_approx} is ready to be described as a tensor network.
We introduce four-leg pure gauge tensors on each plaquette as follows,
\begin{align}
    &T^{(g)}_{x_{g}y_{g}x'_{g}y'_{g}}
    \nonumber\\
    &=
    \begin{cases}
        \displaystyle
        \frac{\sqrt{w_{x_{g}}w_{y_{g}}w_{x'_{g}}w_{y'_{g}}}}{2^{2}}
        \exp\left[-\beta\frac{1-\cos\pi\left(y'_{g}+x_{g}-y_{g}-x'_{g}\right)}
        {1-\left[1-\cos\pi\left(y'_{g}+x_{g}-y_{g}-x'_{g}\right)\right]/\epsilon}\right]
         & {\rm if~admissible} \\
        0 & {\rm otherwise}
    \end{cases}
    ,
\end{align}
\begin{align}
    T^{(\theta)}_{x_{g}y_{g}x'_{g}y'_{g}}
    =
    \exp\left(
        \frac{\im\theta}{2\pi}
        \ln\left[{\rm e}^{\im\pi\left(y'_{g}+x_{g}-y_{g}-x'_{g}\right)}\right]
    \right)
    .
\end{align}
For the Higgs part, we introduce the following hopping matrix,
\begin{align}
    &H_{\tilde{\ell}(n)\tilde{\theta}_{\nu}(n)\tilde{\ell}(n+\hat{\nu})}
    \nonumber\\
    &=
    \frac{\sqrt[4]{w_{\tilde{\ell}(n)}w_{\tilde{\ell}(n+\hat{\nu})}}\e^{(\tilde{\ell}(n)+\tilde{\ell}(n+\hat{\nu}))/4}}{\sqrt{2}}
    \nonumber\\
    &\times\exp\left[
        2\sqrt{\tilde{\ell}(n)\tilde{\ell}(n+\hat{\nu})}
        \cos\pi\tilde{\theta}_{\nu}(n)
        -\frac{M}{4}\left(\tilde{\ell}(n)+\tilde{\ell}(n+\hat{\nu})\right)
        -\frac{\lambda}{4}\left(\tilde{\ell}(n)^{2}+\tilde{\ell}(n+\hat{\nu})^{2}\right)
    \right]
    .
\end{align}
Now, we perform the singular value decomposition (SVD) of the $\nu$-directional hopping matrix, which gives us
\begin{align}
\label{eq:tolerance}
    H_{\tilde{\ell}(n)\tilde{\vartheta}_{\nu}(n)\tilde{\ell}(n+\hat{\nu})}
    \simeq
    \sum_{\alpha=1}^{\chi}
    A_{\tilde{\ell}(n)\tilde{\vartheta}_{\nu}(n)\alpha}
    B_{\tilde{\ell}(n+\hat{\nu})\alpha}
    ,
\end{align}
where $A$ and $B$ are defined by unitary matrices multiplied by the square root of singular values $\sigma_{\alpha}$.
In this study, we choose $\chi$ in Eq.~\eqref{eq:tolerance} such that the singular values satisfying $\sigma_{\alpha}/\sigma_{1}>10^{-7}$ are kept. 
Note that $\sigma_{1}$ is the largest singular value and $\sigma_{\alpha}$ is in the descending order.
We are now allowed to integrate out $\tilde{\ell}(n)$ at each site $n$.
As a result, we can define a six-leg tensor at each lattice site as,
\begin{align}
    T^{(h)}_{x_{h}y_{h}x'_{g}x'_{h}y'_{g}y'_{h}}
    =
    \sum_{\tilde{\ell}(n)}
    A_{\tilde{\ell}(n)y'_{g}x_{h}}
    A_{\tilde{\ell}(n)x'_{g}y_{h}}
    B_{\tilde{\ell}(n)x'_{h}}
    B_{\tilde{\ell}(n)y'_{h}}
    .
\end{align}
Therefore, the tensor network representation for Eq.~\eqref{eq:z_approx} is obtained as
\begin{align}
\label{eq:tn_rep}
    Z(K_{g},K_{h})
    =
    {\rm tTr}\left[\prod_{n}T_{n}\right]
    ,
\end{align}
with the fundamental tensor $T_{n}$ at each site $n$,
\begin{align}
\label{eq:fundamental_tensor}
    (T_{n})_{xyx'y'}
    =
    T^{(g)}_{x_{g}y_{g}x'_{g}y'_{g}}
    T^{(\theta)}_{x_{g}y_{g}x'_{g}y'_{g}}
    T^{(h)}_{x_{h}y_{h}x'_{g}x'_{h}y'_{g}y'_{h}}
    ,
\end{align}
whose bond dimension is $K_{g}\chi $.
Note that the subscripts in the left-hand side of Eq.~\eqref{eq:fundamental_tensor} are defined as $i=(i_{g}i_{h})$ with $i=x,y,x',y'$.

\subsection{Coarse-graining algorithm}

We apply the bond-weighted TRG (BTRG) algorithm~\cite{PhysRevB.105.L060402} to approximately compute the path integral in Eq.~\eqref{eq:tn_rep}.
BTRG allows us to approximately carry out the contractions among $2^{q}$ local tensors within the $q$ times of coarse-graining.
Since each local tensor in Eq.~\eqref{eq:fundamental_tensor} is defined on each lattice site, $q$ relates to the volume $V$ via $V=2^{q}$ and the linear system size $L$ via $L=2^{q/2}$.
For the algorithmic details, see Ref.~\cite{PhysRevB.105.L060402}.
\footnote{
Since the model is defined on a square lattice, we always set the hyperparameter in the BTRG algorithm as $k=-0.5$~\cite{PhysRevB.105.L060402,Akiyama:2022pse}.
}

This algorithm improves the accuracy of the original Levin-Nave TRG at the same bond dimension.
Remarkably, the computational cost of BTRG is completely the same as the Levin-Nave TRG.
In benchmarking by the 2$d$ Ising model in Ref.~\cite{PhysRevB.105.L060402}, the BTRG shows better performance not only than the Levin-Nave TRG but also than the higher-order TRG~\cite{PhysRevB.86.045139}, one of the most commonly used TRG algorithms, whose computational cost is greater than the Levin-Nave TRG as well as BTRG.
The essence of the BTRG is to introduce a weight on each edge of the tensor network.
These weights mimic the effect of the environment, which is not taken into account in the original Levin-Nave TRG.
Therefore, the BTRG can be regarded as a variant of the second renormalization group (SRG) algorithms~\cite{Xie:2009zzd,PhysRevB.81.174411}, but without any backward iteration to update the environment tensors as in the conventional SRG algorithms.
Moreover, there is no variational determination of local tensors as in the tensor network renormalization (TNR)~\cite{PhysRevLett.115.180405,PhysRevB.95.045117} and loop-TNR~\cite{PhysRevLett.118.110504}.
Since such variational calculations are time-consuming, it is usually difficult to increase the bond dimension in these algorithms, and their application to the lattice model with continuous degrees of freedom might be limited.

\section{Numerical results} 
\label{sec:results}

In the following, we always set $\beta=3.0$ and the positive constant $\epsilon$ in Eq.~\eqref{eq:s_luscher} as $\epsilon=1.0$.
For the gauge-Higgs model, the quartic coupling is fixed as $\lambda=0.5$.
Note that the cutoff effects from the finite lattice spacing of the L{\"u}scher gauge action and standard Wilson gauge action are different even at the same inverse gauge coupling $\beta$.
With the same value of $\beta$, the L{\"u}scher gauge action is expected to be closer to the continuum limit than the Wilson action.
See Appendix~\ref{app:luscher_vs_wilson} for the comparison between the L{\"u}scher and Wilson gauge actions.

\subsection{Pure U(1) gauge theory} 
\label{subsec:pure}

We start by studying the (1+1)$d$ pure gauge theory with a $\theta$ term to validate our tensor network formulation.
In this case, the local tensor in Eq.~\eqref{eq:fundamental_tensor} is defined without $T^{(h)}$ and the bond dimension in Eq.~\eqref{eq:tn_rep} is equal to $K_{g}$.

\begin{figure}[htbp]
    \centering
    \includegraphics[width=0.8\hsize]{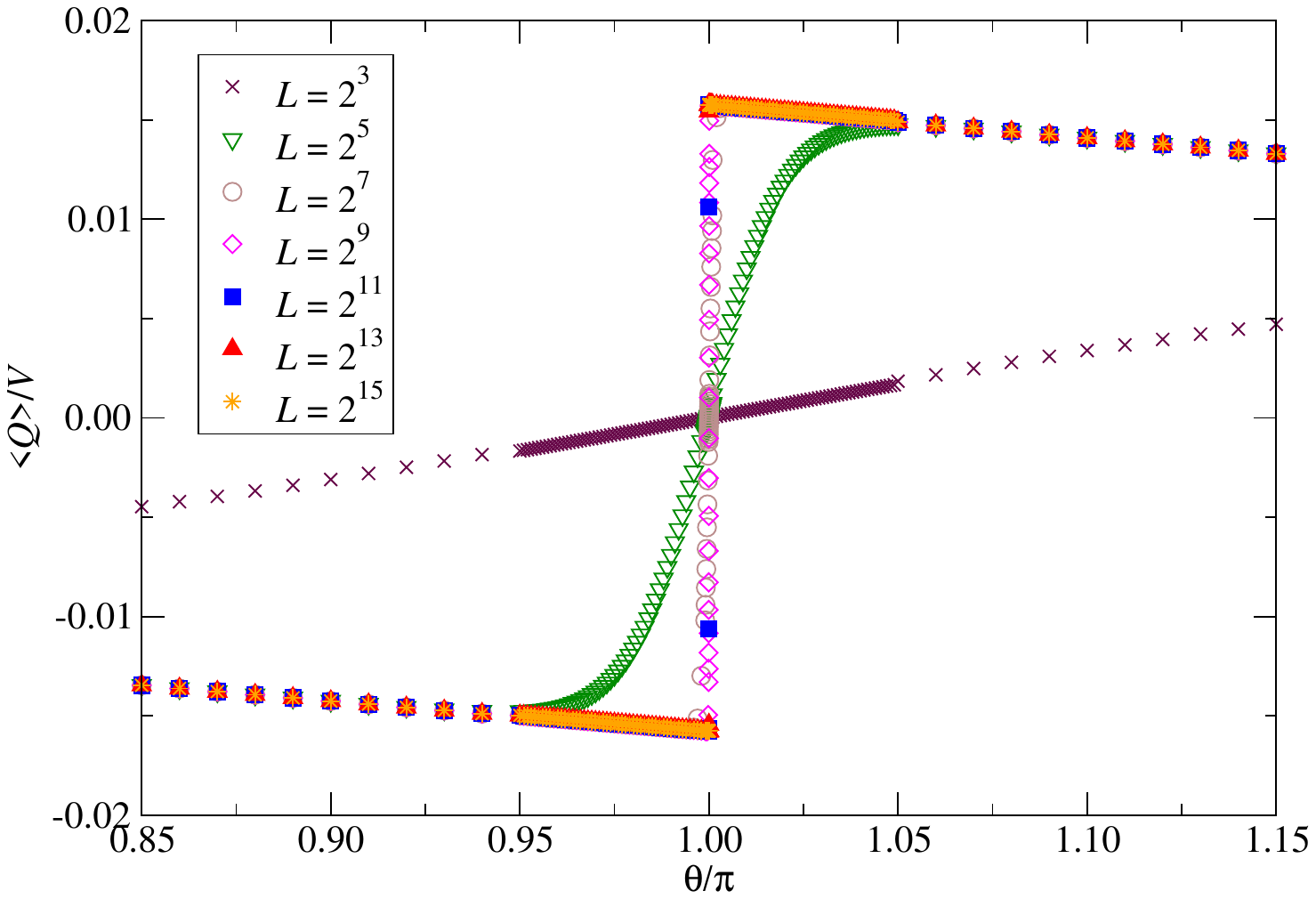}
    \caption{
        Topological charge density for the pure U(1) gauge theory as a function of $\theta/\pi$ at $\beta=3.0$ with $D=K_{g}=30$.
    }
    \label{fig:pure_gauge_top_charge_k30}
\end{figure}

At $\theta=\pi$, the theory is expected to undergo the first-order transition.
We calculate the topological charge density $\langle Q\rangle/V$, which is defined by
\begin{align}
\label{eq:def_top}
    \frac{\langle Q\rangle}{V}
    =
    -\frac{\im}{V}\frac{\partial \ln Z}{\partial \theta}
    .
\end{align}
Figure~\ref{fig:pure_gauge_top_charge_k30} shows the volume dependence of $\langle Q\rangle/V$ at $\beta=3.0$ as a function of $\theta$. 
We set $D=K_{g}=30$ and employed the numerical differentiation to evaluate Eq.~\eqref{eq:def_top}.
As the volume increases, the topological charge density gradually becomes discontinuous and we observe a clear jump at $\theta=\pi$ with $L\ge2^{13}$.
This is a signal of the first-order transition at $\theta=\pi$.

\begin{figure}[htbp]
    \centering
    \includegraphics[width=0.8\hsize]{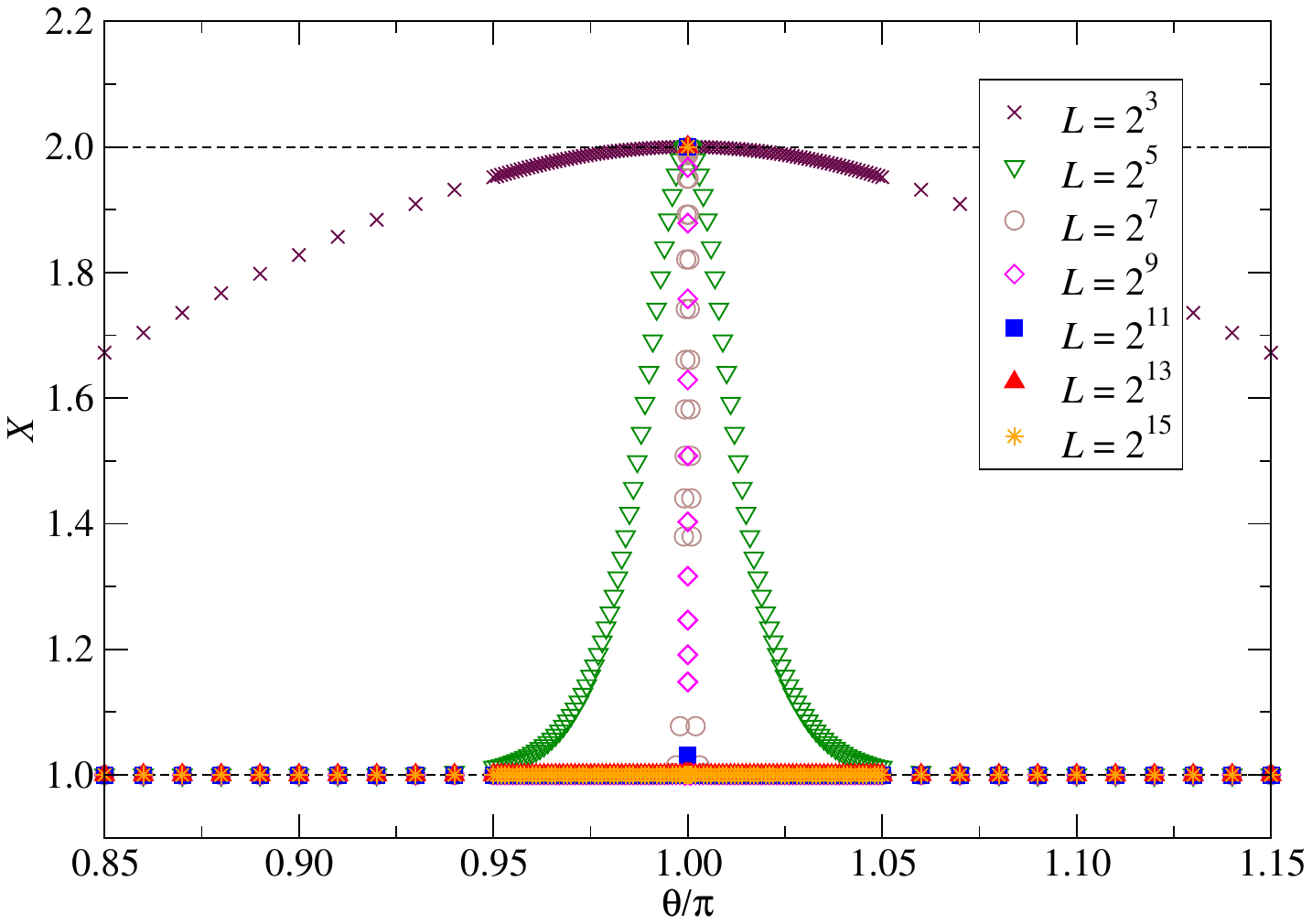}
    \caption{
        The ground state degeneracy $X$ of Eq.~(\ref{eq:x}) for the pure U(1) gauge theory as a function of $\theta/\pi$ at $\beta=3.0$ with $D=K_{g}=30$.
    }
    \label{fig:pure_gauge_x_k30}
\end{figure}

Taking advantage of the TRG method, we also compute the ground state degeneracy by introducing 
\begin{align}
\label{eq:x}
    X
    =
    \frac{\left(\Tr A\right)^{2}}{\Tr\left(A^{2}\right)}
    ,
\end{align}
following Ref.~\cite{PhysRevB.80.155131}.
After sufficient times of coarse-graining, this quantity counts the ground state degeneracy.
In Eq.~\eqref{eq:x}, $A$ is a transfer matrix defined from the local tensor via
\begin{align}
\label{eq:def_transfer_mat}
    A_{yy'}
    =
    \sum_{x}T_{xyxy'}
    .
\end{align}
Figure~\ref{fig:pure_gauge_x_k30} shows $X$ in Eq.~\eqref{eq:x} as a function of $\theta$.
With sufficiently large volume, $X=2$ is realized only at $\theta=\pi$.
Therefore, the current TRG computation successfully reproduces the spontaneous $\mathds{Z}_{2}$ symmetry breaking at $\theta=\pi$ as expected.

\begin{figure}[htbp]
    \centering
    \includegraphics[width=0.8\hsize]{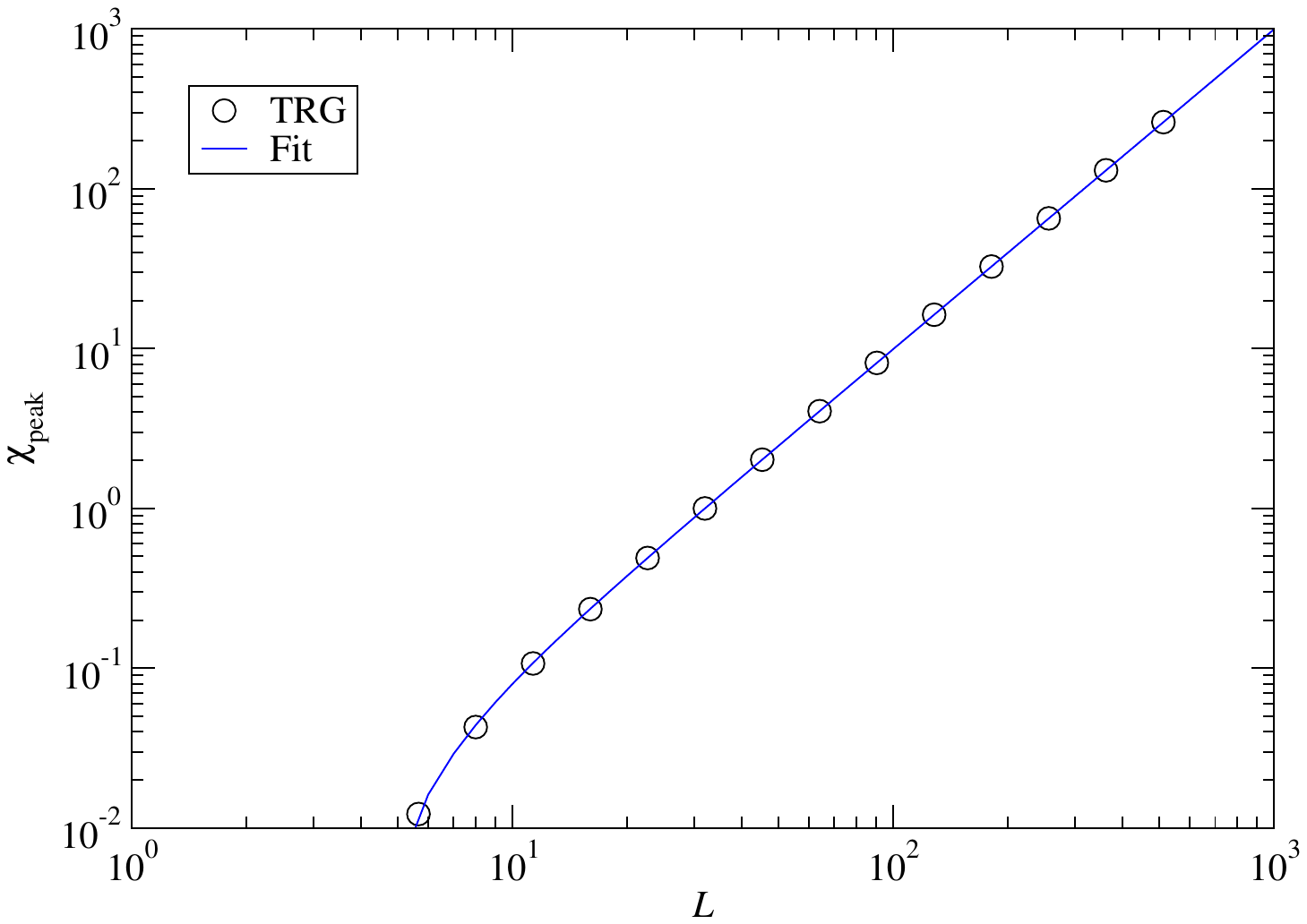}
    \caption{
    System size dependence of $\chi_{\rm peak}$ for the pure U(1) gauge theory at $\beta=3.0$ with $D=K_{g}=30$.
    }
    \label{fig:pure_gauge_chi_max_k30}
\end{figure}

Additionally, we apply the finite-size scaling analysis for the topological susceptibility, which is given by
\begin{align}
\label{eq:chi_top}
    \chi_{Q}
    =
    \frac{\partial}{\partial\theta}
    \frac{\langle Q\rangle}{V}
    .
\end{align}
We use the numerical differentiation with the $O(\Delta^{4})$ accuracy, setting $\Delta=\pi\times10^{-5}$ around $\theta=\pi$, to evaluate Eq.~\eqref{eq:chi_top}.
Since the peak of the topological susceptibility always appears at $\theta=\pi$, we examine the system size dependence of the values of $\chi_{Q}$ at $\theta=\pi$.
Let $\chi_{\rm peak}(L)$ be the value of $\chi_{Q}$ at $\theta=\pi$ with the linear system size $L$.
Using the following ansatz,
\begin{align}
    \chi_{\rm peak}(L)
    =
    c_{0}
    +
    c_{1}L^{p}
    ,
\end{align}
the data of $\chi_{\rm peak}(L)$ are fitted as shown in Figure~\ref{fig:pure_gauge_chi_max_k30}.
The fit is performed on the data over $L\in[4,256\sqrt{2}]$ and we obtain $p=2.00001(6)$, $c_{0}=-0.0196(6)$, and $c_{1}=0.0009944(3)$.
Therefore, $\chi_{\rm peak}(L)$ is proportional to the volume and this is another indication of the first-order phase transition.
Although we employed the finite bond dimension $D$ and cutoff $K_{g}$, the current computation correctly reproduces the first-order phase transition at $\theta=\pi$.
We have also tried the same analysis with $D=K_{g}=20$, which results in $p=2.00000(5)$, $c_{0}=-0.0197(5)$, and $c_{1}=0.0009973(3)$.
Therefore, $D, K_{g}\ge20$ seems sufficiently large to investigate the phase transition at $\theta=\pi$ for the pure U(1) gauge theory at $\beta=3.0$.

\subsection{The U(1) gauge-Higgs model} 
\label{subsec:gaugehiggs}

\begin{figure}[htbp]
    \centering
    \includegraphics[width=0.8\hsize]{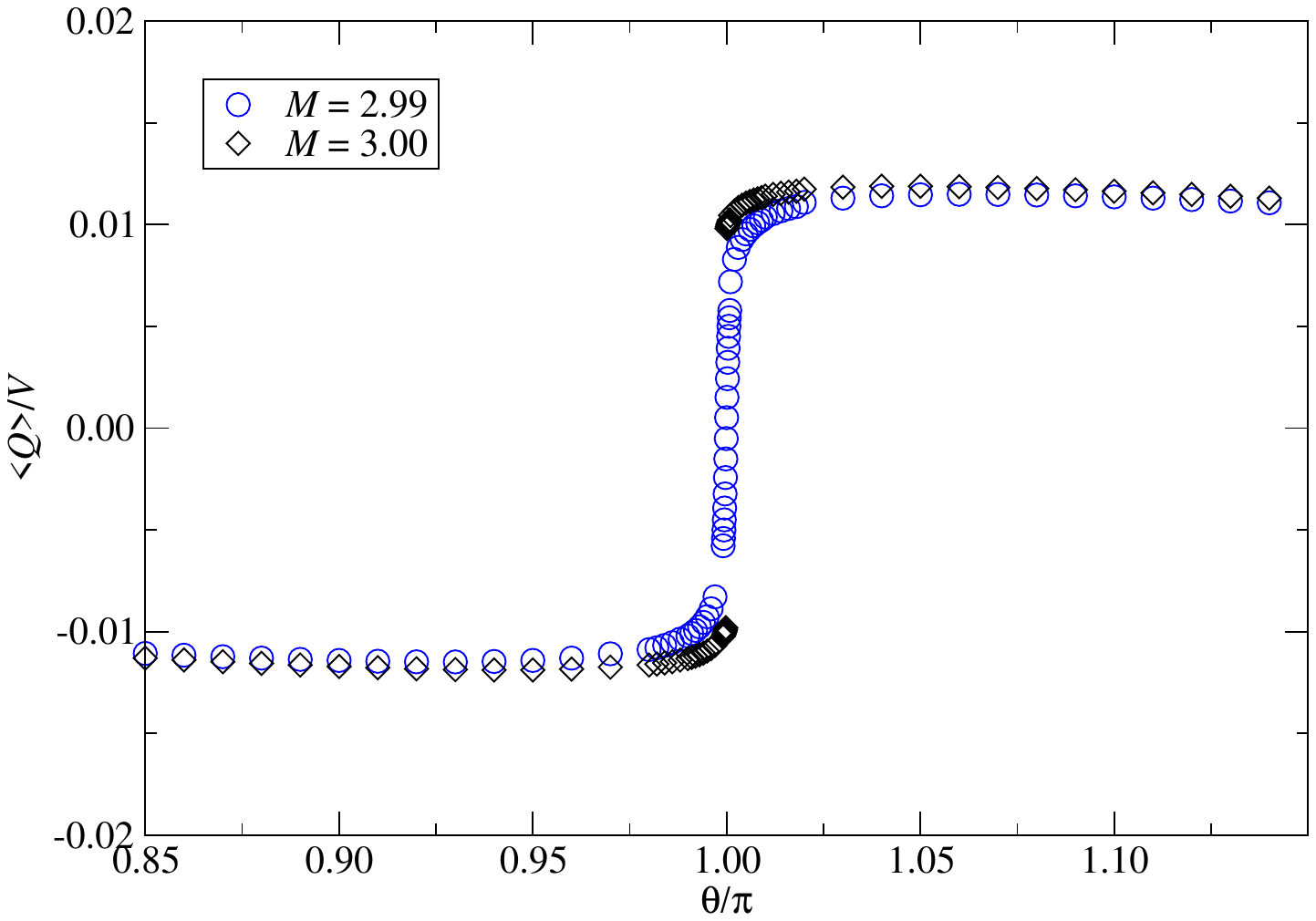}
    \caption{
        Topological charge density in the thermodynamic limit for the U(1) gauge-Higgs model as a function of $\theta/\pi$ at $\beta=3.0$ and $\lambda=0.5$ with $K_{g}=20$, $K_{h}=20$, and $D=160$.  
        The circle and diamond denote $M=2.99$ and $M=3.00$, respectively.
    }
  	\label{fig:gauge_higgs_top_charge}
\end{figure}

We now investigate the critical behavior of the U(1) gauge-Higgs model at $\theta=\pi$.
Figure~\ref{fig:gauge_higgs_top_charge} shows the thermodynamic topological charge density at $M=2.99$ and $M=3.00$, where we set $K_{g}=20$, $K_{h}=20$ which result in $\chi=8$ according to Eq.~\eqref{eq:tolerance}.
We also set $D=K_{g}\chi=160$.
As we will see, our choice of $D=K_{g}\chi=160$ is sufficiently large to determine the universality class at the critical endpoint.
The behavior of topological charge density is highly different between these two mass parameters; $\langle Q\rangle/V$ varies continuously around $\theta=\pi$ at $M=2.99$ but the discontinuity appears at $\theta=\pi$ when $M=3.00$.
Therefore, the critical mass $M_{\rm c}$ should exist in the range of $2.99\le M_{\rm c}\le3.00$.
\begin{figure}[htbp]
    \centering
    \includegraphics[width=0.8\hsize]{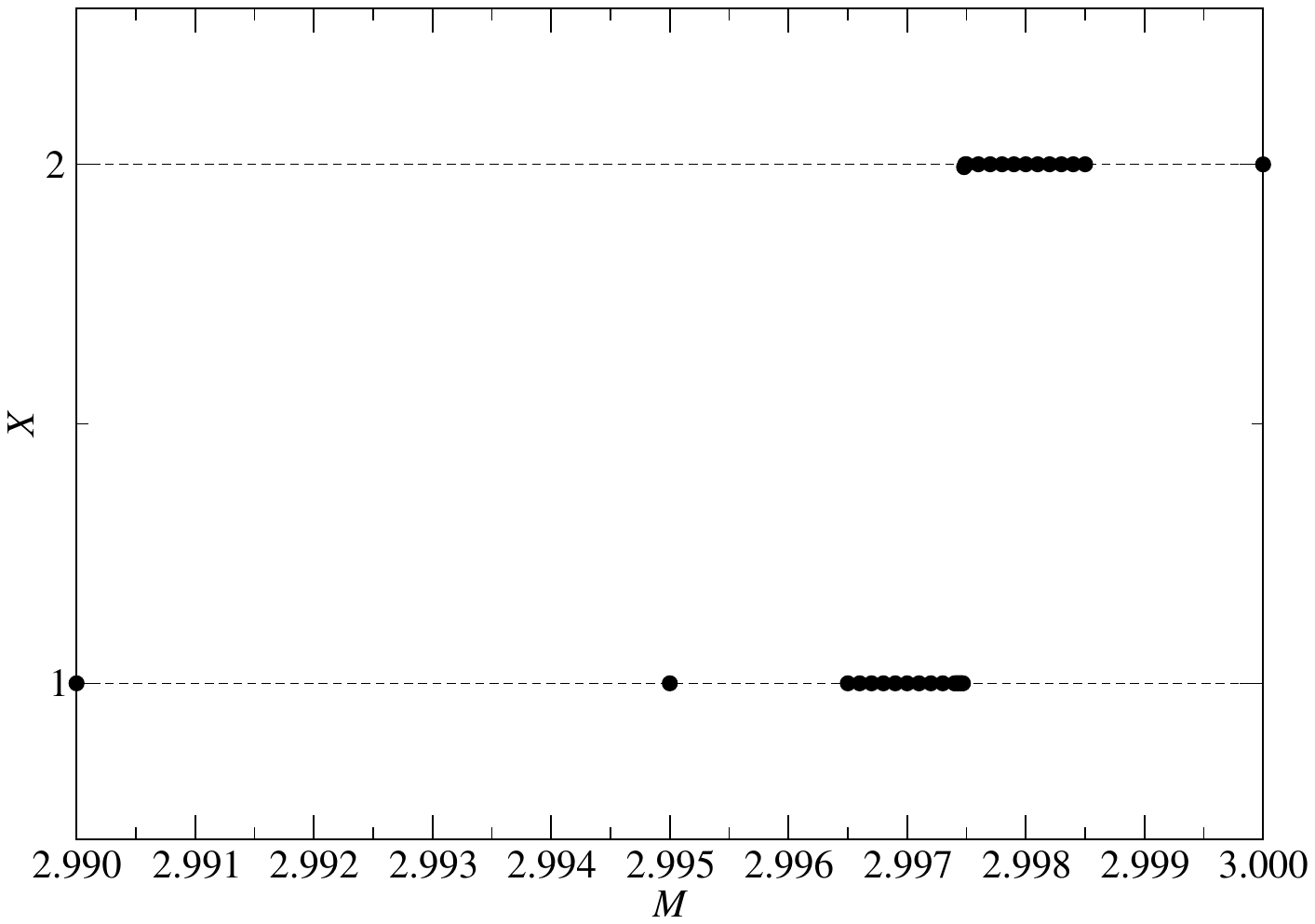}
    \caption{
        The ground state degeneracy $X$ computed on $V=2^{40}$ as a function of $M$.
    }
    \label{fig:gauge_higgs_x}
\end{figure}
To locate the critical point $M_{\rm c}$ more precisely, we use the ground state degeneracy $X$ in Eq.~\eqref{eq:x}.
When $M>M_{\rm c}$, we should have $X=2$ due to the $\mathds{Z}_{2}$ charge symmetry breaking and with $M<M_{\rm c}$, we must have $X=1$.
Figure~\ref{fig:gauge_higgs_x} shows the degeneracy $X$ as a function of $M$, which results in $2.99747\le M_{\rm c}\le2.99748$.

\begin{figure}[htbp]
    \centering
    \includegraphics[width=0.8\hsize]{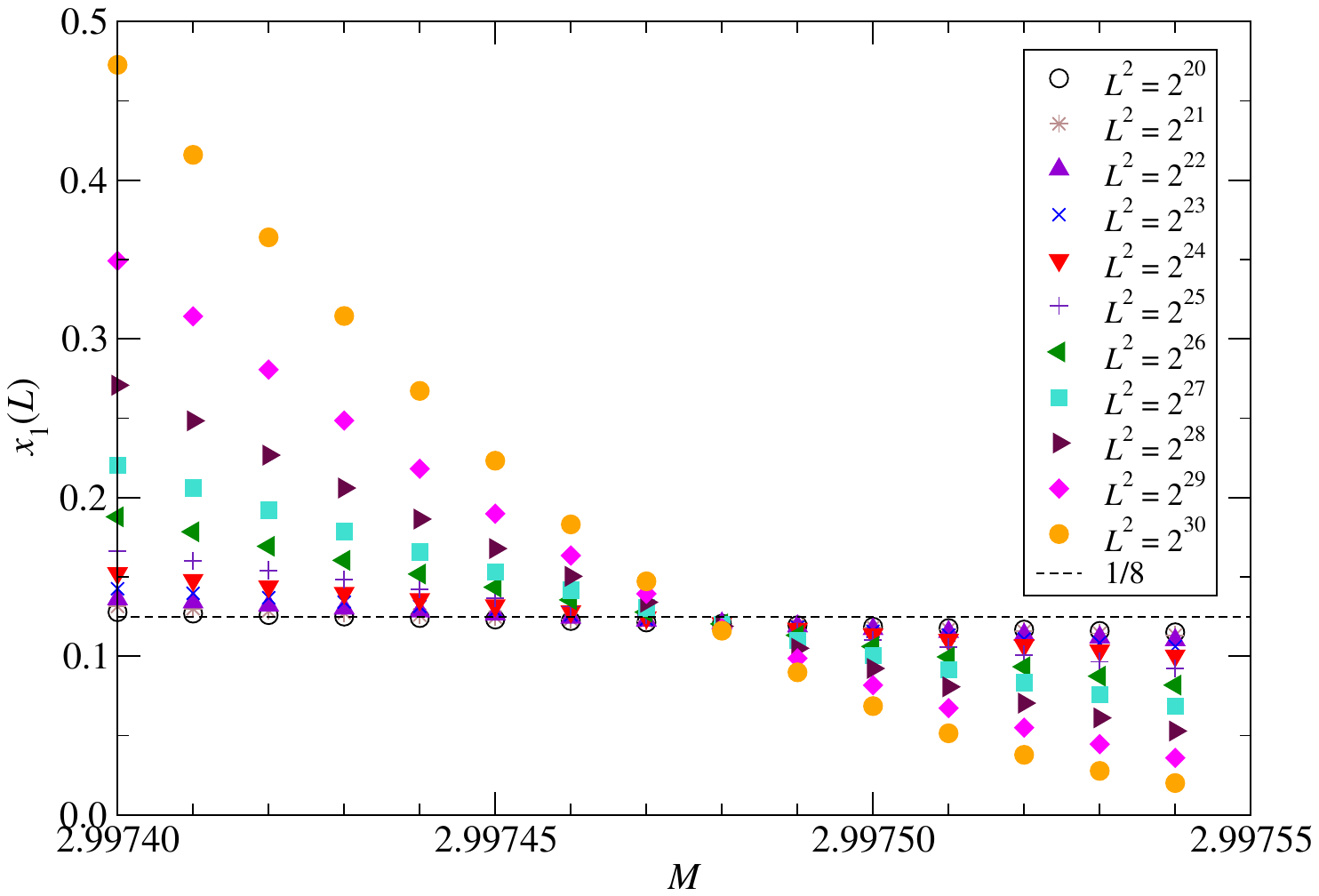}
    \caption{
        The system-size dependence of the scaling dimension $x_{1}(L)$.
        The dashed line denotes the theoretical value of the $2d$ Ising universality class, $x_{1}=1/8$.
    }
    \label{fig:scaling_dimension_x_sigma}
\end{figure}

\begin{figure}[htbp]
    \centering
    \includegraphics[width=0.8\hsize]{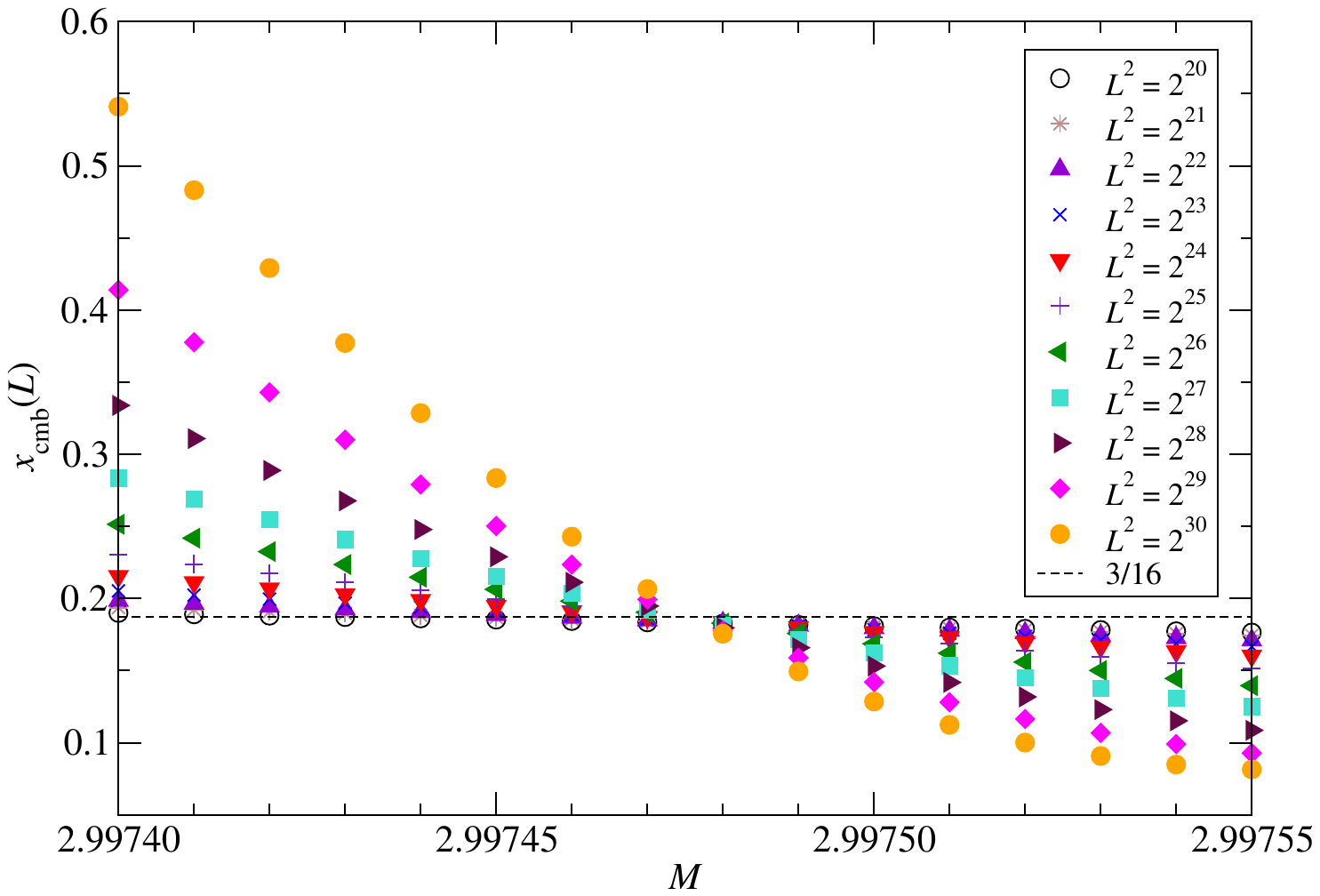}
    \caption{
        The system-size dependence of the scaling dimension $x_{\rm cmb}(L)$.
        The dashed line denotes the theoretical value of the $2d$ Ising universality class, $x_{\rm cmb}=3/16$.
    }
    \label{fig:scaling_dimension_x_cmb}
\end{figure}

To identify the universality class, we analyze the eigenvalue spectrum of the transfer matrix $A$ defined in Eq.~\eqref{eq:def_transfer_mat}.
Note that the analysis on the transfer matrix obtained from the TRG has been employed in investigating several spin models~\cite{PhysRevB.104.165132,PhysRevB.80.155131,Huang:2023vwp,PhysRevB.108.024413,Az-zahra:2024gqr} and the lattice Schwinger model~\cite{Shimizu:2017onf}.
For a comprehensive overview, see Ref.~\cite{ueda2024renormalizationgroupflowfixedpoint}.
Although we are dealing with continuous gauge and scalar degrees of freedom, a similar analysis should be possible if our discretization scheme based on the Gauss quadrature rules is working.
Particularly, we employ the tensor-network-based level spectroscopy proposed in Refs.~\cite{PhysRevB.104.165132,PhysRevB.108.024413}.
Let $\lambda_{n}~(n=0,1,\cdots)$ be an $n$-th eigenvalue of the transfer matrix $A$.
Assuming that these eigenvalues are in descending order, we can obtain the scaling dimension $x_{n}(L)$ at the finite system size $L$ via
\begin{align}
    x_{n}(L)
    =
    \frac{1}{2\pi}\ln\frac{\lambda_{0}(L)}{\lambda_{n}(L)}.
\end{align}
When these quantities are computed at the criticality and on a sufficiently large volume, they give us the scaling dimension of the conformal field theory (CFT).
Figure~\ref{fig:scaling_dimension_x_sigma} shows $x_{1}(L)$ as a function of the Higgs mass $M$.
The volume independence can be observed at $M\sim2.99748$ with $x_{1}=1/8$, which is consistent with the 2$d$ Ising universality class.
From now on, we assume the critical phenomenon is in the 2$d$ Ising universality class.
Following Ref.~\cite{PhysRevB.108.024413}, we consider a combined scaling dimension,
\begin{align}
    x_{\rm cmb}(L)
    =
    x_{1}(L)+\frac{1}{16}x_{2}(L),
\end{align}
which removes the effect of the leading irrelevant perturbation associated with the scaling dimension 4.
Since $x_{2}=1$ in the 2$d$ Ising universality class, $x_{\rm cmb}=3/16$ is expected at the criticality.
Figure~\ref{fig:scaling_dimension_x_cmb} shows $x_{\rm cmb}(L)$ against the Higgs mass $M$ which is in agreement with this expectation.
Note that $x_{\rm cmb}(L)$ shown in Figure~\ref{fig:scaling_dimension_x_cmb} also agrees with the previous estimation by $X$: $2.99747\le M_{\rm c}\le2.99748$.
Moreover, we apply a scheme to determine the critical point following Ref.~\cite{PhysRevB.108.024413}, which assumes the phase transition is in the $2d$ Ising universality class.
We first choose two mass parameters $M^{(+)}$ and $M^{(-)}$ such that $M^{(-)}\le M_{\rm c}\le M^{(+)}$ and compute $\delta x_{\rm cmb}(L)=x_{\rm cmb}(L)-3/16$.
When $M\le M_{\rm c}$ ($M\ge M_{\rm c}$), $\delta x_{\rm cmb}(L)$ should increase (decrease) by enlarging the system size as shown in Fig.~\ref{fig:scaling_dimension_x_cmb}.
Secondly, we perform linear interpolations of $\delta x_{\rm cmb}(L)$ between $M^{(+)}$ and $M^{(-)}$, and find a crossing point $M^{*}(L)$ of two lines with the system size $L$ and $\sqrt{2}L$.
Then, $M_{\rm c}$ is obtained via $M^{*}(L)=M_{\rm c}+aL$, where $M_{\rm c}$ and $a$ are the parameters determined by the numerical fit.
The fit results in $M_{\rm c}=2.997480(2)$, where we set $M^{(-)}=2.99747$, $M^{(+)}=2.99749$ using the data with $L\in[2^{10},2^{14}\sqrt{2}]$.
Therefore, the tensor-network-based level spectroscopy confirms the critical point estimated by the ground state degeneracy.
We also investigate the finite-size correction for the free energy in the thermodynamic limit via
\begin{align}
\label{eq:central_charge}
    \frac{1}{L^{2}}\ln\lambda_{0}
    =
    -f_{\infty}
    +
    \frac{\pi c}{6L^{2}}
    ,
\end{align}
where $f_{\infty}$ is the thermodynamic free energy and $c$ is the central charge of the CFT at criticality.
Using Eq.~\eqref{eq:central_charge} as a fitting ansatz for $\ln\lambda_{0}/L^{2}$, we can determine the central charge.
As a representative point, we choose $M=2.99748$ and the fit using the data with $L\in[2^{10},2^{15}]$ results in $c=0.50(7)$, which is consistent with $c=1/2$, as expected.


We finally remark on the location of $M_{\rm c}$ which slightly depends on the algorithmic parameters, $K_{g}$, $K_{h}$, and $D=K_{g}\chi$.
Table~\ref{tab:comparison_mc} shows the critical endpoint $M_{\rm c}$ estimated by the scheme in Ref.~\cite{PhysRevB.108.024413}.
In Table~\ref{tab:comparison_mc}, the error originates from the fit based on $M^{*}(L)=M_{\rm c}+aL$.
We find that these estimations are comparable with $M_{\rm c}=2.989(2)$ obtained in Ref.~\cite{Gattringer:2018dlw}, where the gauge-field Boltzmann weight is given by the Villain form and the Monte Carlo simulation is performed based on the dual representation.
Note that $M_{\rm c}=2.989(2)$ in Ref.~\cite{Gattringer:2018dlw} is obtained at $\beta=3.0$ and $\lambda=0.5$.

\begin{table}[htbp]
    \caption{
        Comparison of the critical endpoint $M_{\rm c}$ against the algorithmic parameters.
    }
    \begin{center}
        \begin{tabular}{ccccc} \hline
            $K_{g}$ & $K_{h}$ & $\chi$ & $D$ & $M_{\rm c}$ \\ \hline
            24 & 20 & 8 & 192 & 2.9982886(1)\\
            22 & 20 & 8 & 176 & 2.9998263(13)\\
            20 & 20 & 8 & 160 & 2.9974765(14)\\
            24 & 10 & 6 & 144 & 2.9929635(1)\\
            22 & 10 & 6 & 132 & 2.9945222(9)\\
            20 & 10 & 7 & 140 & 2.9921698(6)\\ \hline
        \end{tabular}
    \end{center}
    \label{tab:comparison_mc}
\end{table}


\section{Summary and outlook} 
\label{sec:summary}

We have analyzed the phase structure of the (1+1)$d$ U(1) gauge-Higgs model with a $\theta$ term, whose gauge action is constructed with L{\"u}scher's admissibility condition. 
Although the model suffers from a complex action problem and topological freezing within the Monte Carlo simulation, the TRG approach allows us to deal with the model successfully.
We have observed the first-order phase transition at $\theta=\pi$ with sufficiently large lattice Higgs mass $M$, including the pure gauge theory, with the finite gauge coupling $\beta$ and the quartic coupling $\lambda$. 
We have determined the critical endpoint and its universality class from the numerical analysis of the transfer matrix, which can be directly obtained from the TRG computation.
Employing the tensor-network-based level spectroscopy, we have confirmed that the scaling dimensions are consistent with the 2$d$ Ising universality class.

All these results show that the TRG is a promising approach to deal with the lattice gauge theory with L{\"u}scher's admissibility condition.
We emphasize that one can easily combine the L{\"u}scher gauge action with fermions and extend the theory for the higher dimensions within the tensor network formulation.
It would be an interesting future work to investigate the Schwinger model with a $\theta$ term under L{\"u}scher's admissibility condition.

\begin{acknowledgments}
Numerical calculation for the present work was carried out with SQUID at the Cybermedia Center, Osaka University (Project ID: hp240012).
We also used the supercomputer Fugaku provided by RIKEN through the HPCI System Research Project (Project ID: hp230247) and computational resources of Wisteria/BDEC-01 and Cygnus and Pegasus under the Multidisciplinary Cooperative Research Program of Center for Computational Sciences, University of Tsukuba. 
This work is supported by Grants-in-Aid for Scientific Research from the Ministry of Education, Culture, Sports, Science and Technology (MEXT) (Nos. 24H00214, 24H00940). 
SA acknowledges the support from the Endowed Project for Quantum Software Research and Education, the University of Tokyo~\cite{qsw}, JSPS KAKENHI Grant Number JP23K13096, and the Center of Innovations for Sustainable Quantum AI (JST Grant Number JPMJPF2221).
\end{acknowledgments}

\appendix

\section{Comparison with the Wilson gauge action}
\label{app:luscher_vs_wilson}

Employing the pure U(1) gauge theory with a $\theta$ term, we make a comparison between the L{\"u}scher gauge action in Eq.~\eqref{eq:gauge_luscher} and the Wilson gauge action,
\begin{align}
    S_{g}=
    \sum_{n}
    \left(
        1-{\rm Re}P_{12}(n)
    \right)
    .
\end{align}
Here, we consider a $\theta$ term defined by
\begin{align}
\label{eq:theta_sin}
    S_{\theta}
    =
    \frac{\im\theta}{2\pi}
    \sum_{n}{\rm Im} P_{12}(n)
    ,
\end{align}
instead of Eq.~\eqref{eq:theta_log}.
In the continuum limit, Eq.~\eqref{eq:theta_sin} reproduces
\begin{align}
    S_{\theta}
    =
    \frac{\im\theta}{2\pi}
    \int{\rm d}^{2}x F_{12}(x)
    ,
\end{align}
which is a topological term of the continuum theory and $F_{12}(x)$ is the field strength in two dimensions.
The continuum limit is defined by $\beta\to\infty$ and $V\to\infty$ with fixing $\beta/V$.
The $\theta$ term in Eq.~\eqref{eq:theta_sin} reproduces the $2\pi$ periodicity of observables as a function of $\theta$ with sufficiently large $\beta$ and $V$~\cite{Gattringer:2015baa}.

\begin{figure}[htbp]
    \centering
    \text{\small (a) $(\beta,V)=(1.6,2^{4})$}\\
    \begin{tabular}{cc}
        \begin{minipage}[t]{0.43\hsize}
            \centering
            \includegraphics[width=1.0\hsize] {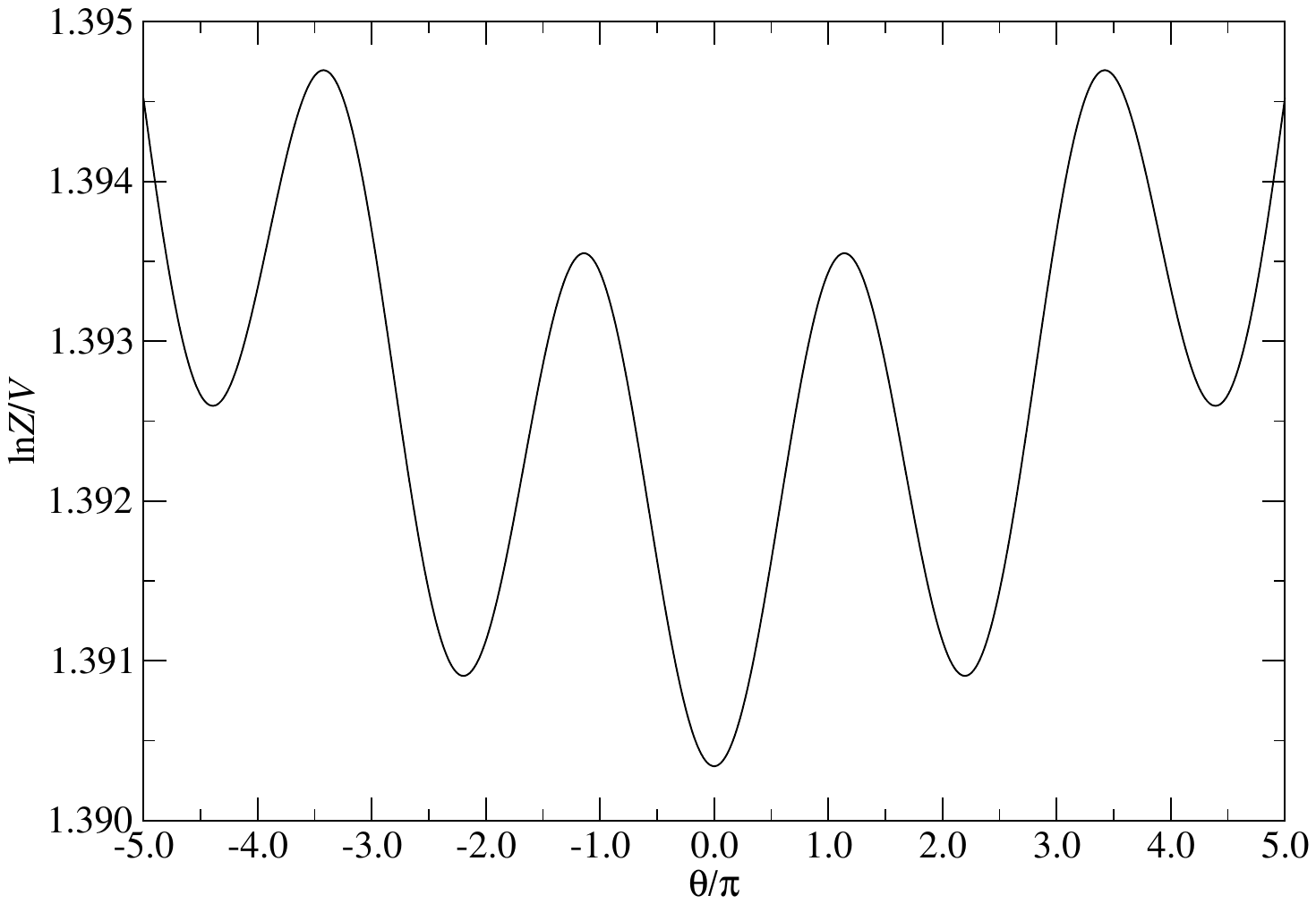} 
        \end{minipage} &
        \begin{minipage}[t]{0.43\hsize}
            \centering
            \includegraphics[width=1.0\hsize]{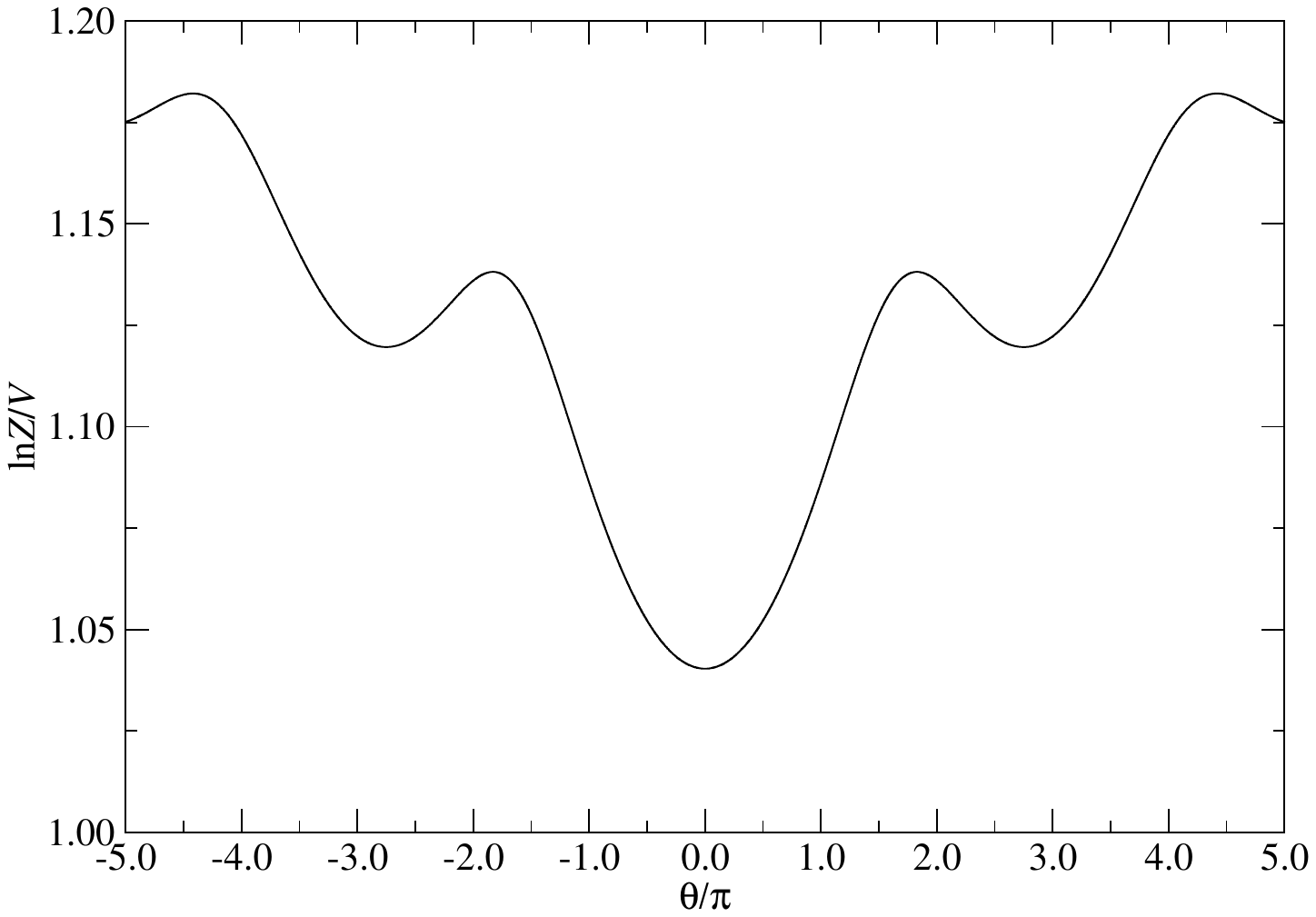}
        \end{minipage} 
    \end{tabular}
    \centering
    \text{\small (b) $(\beta,V)=(3.2,2^{5})$}\\
    \begin{tabular}{cc}
        \begin{minipage}[t]{0.43\hsize}
            \centering
            \includegraphics[width=1.0\hsize] {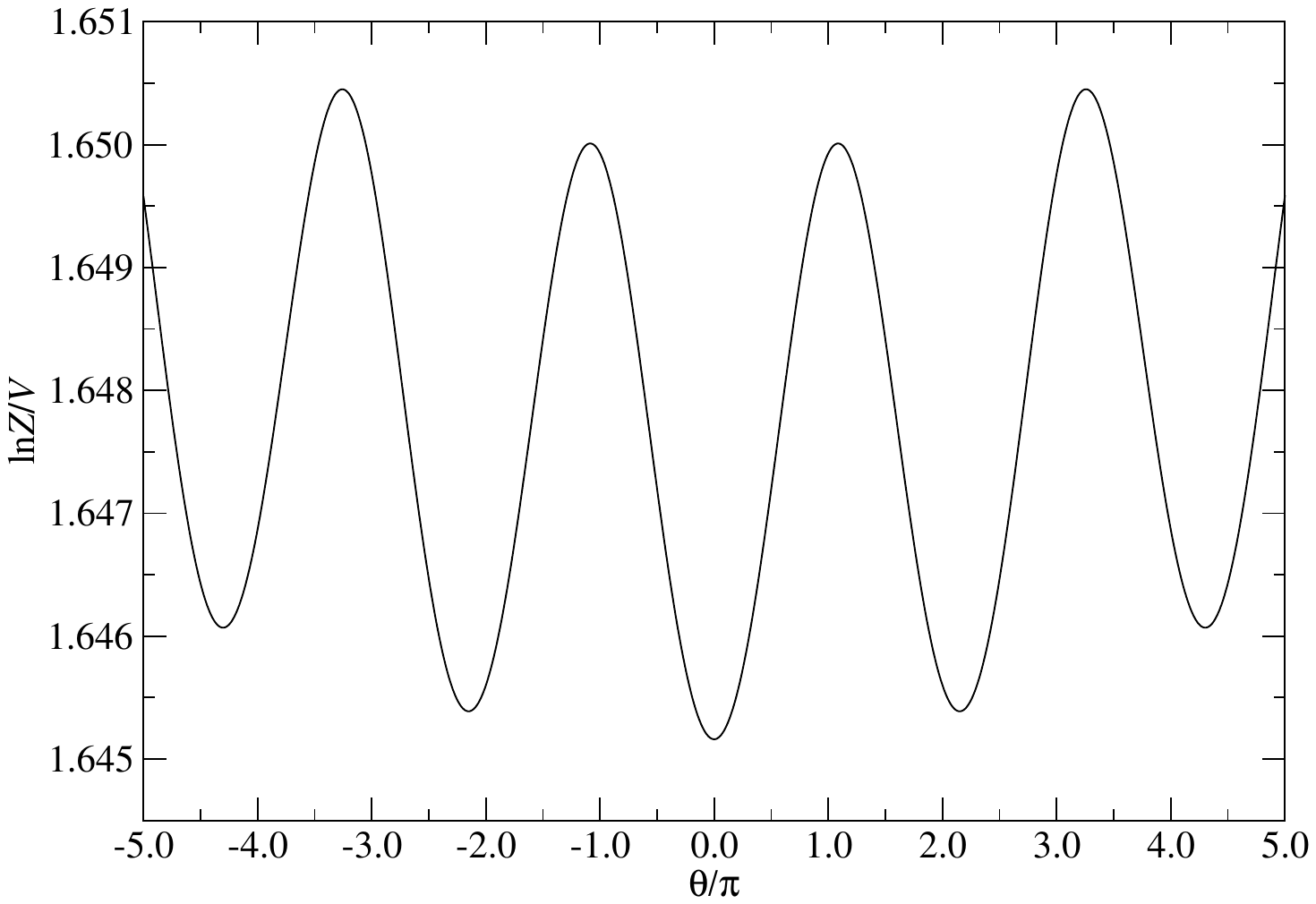} 
        \end{minipage} &
        \begin{minipage}[t]{0.43\hsize}
            \centering
            \includegraphics[width=1.0\hsize]{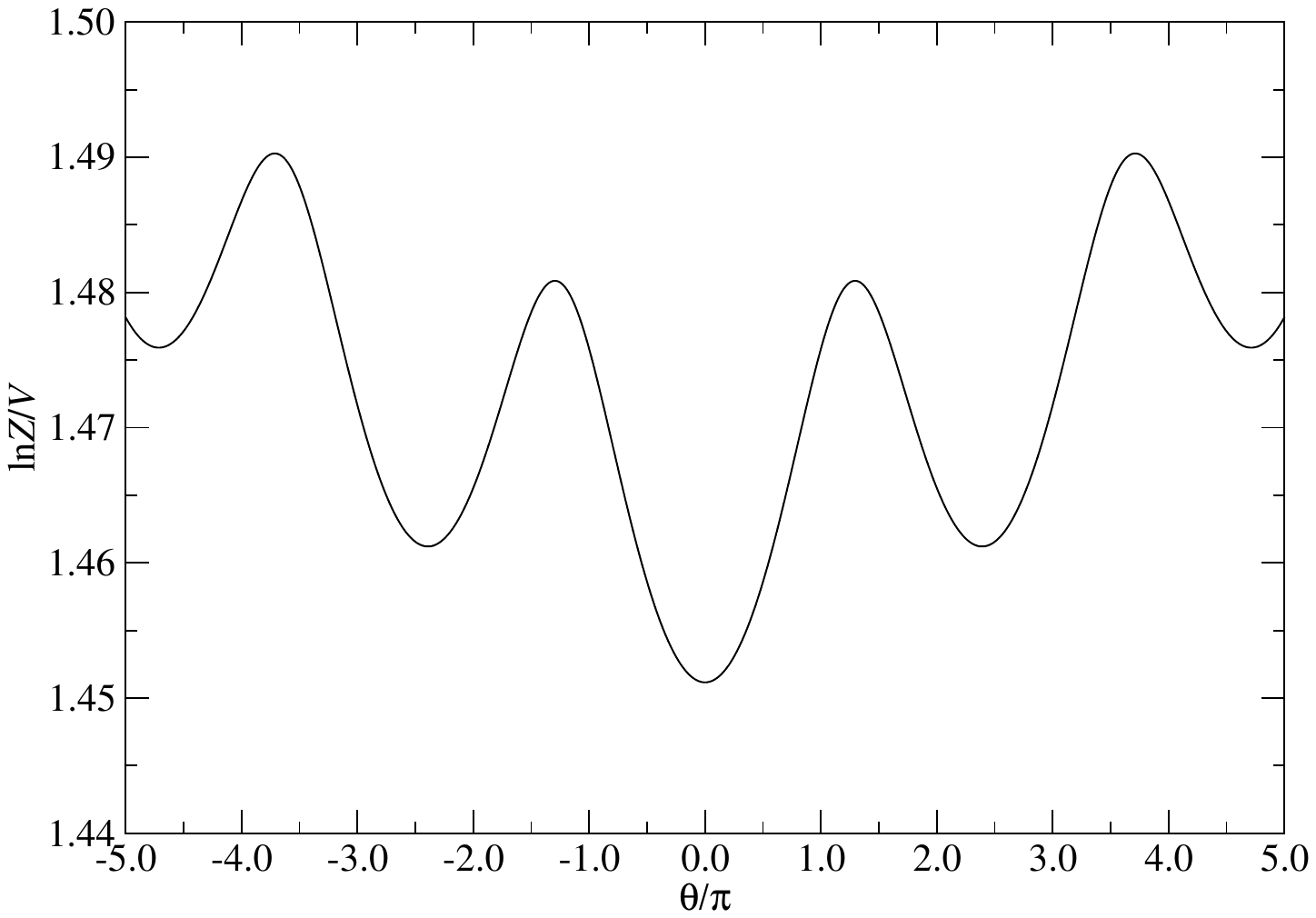}
        \end{minipage} 
    \end{tabular}
    \centering
    \text{\small (c) $(\beta,V)=(6.4,2^{6})$}\\
    \begin{tabular}{cc}
        \begin{minipage}[t]{0.43\hsize}
            \centering
            \includegraphics[width=1.0\hsize] {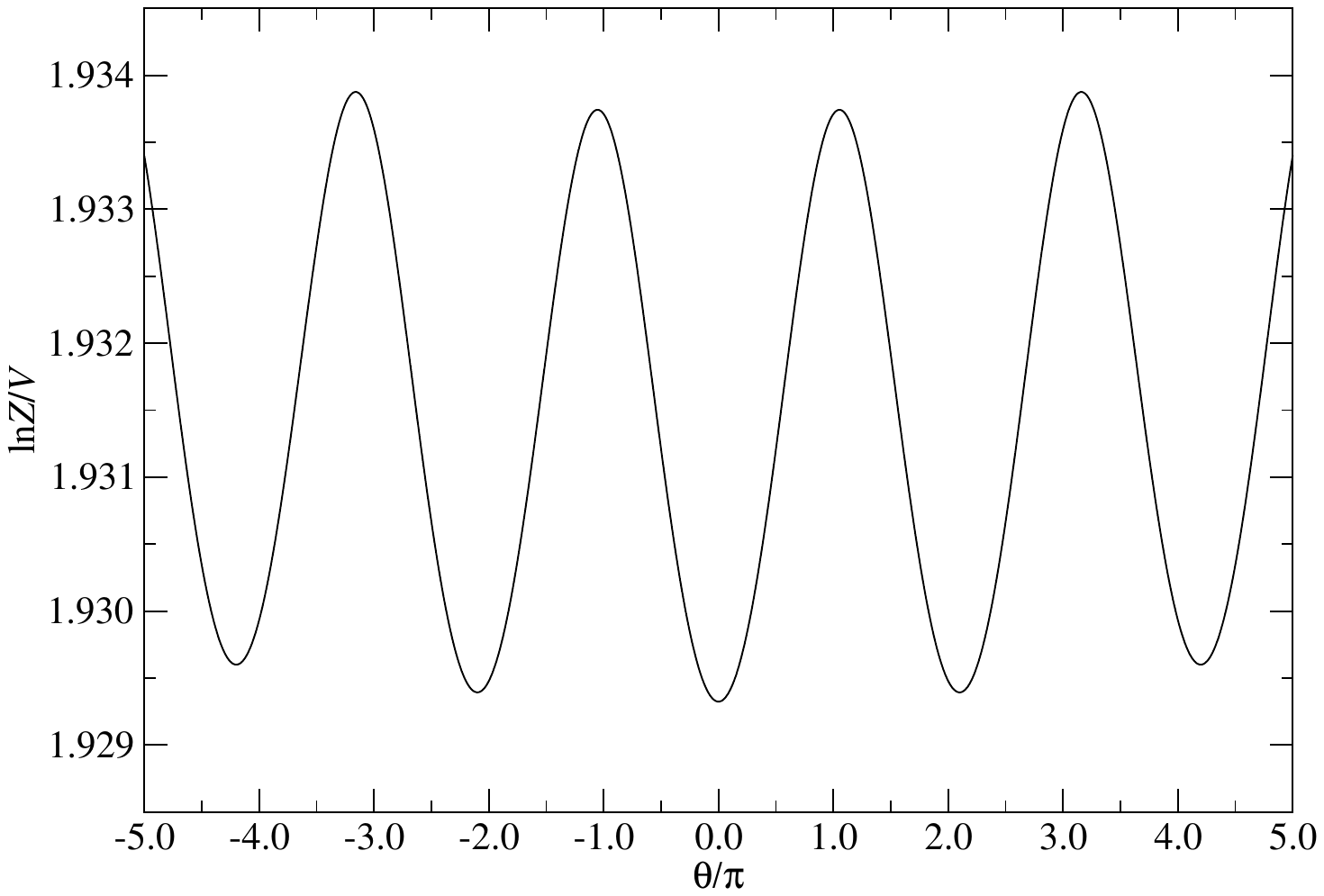} 
        \end{minipage} &
        \begin{minipage}[t]{0.43\hsize}
            \centering
            \includegraphics[width=1.0\hsize]{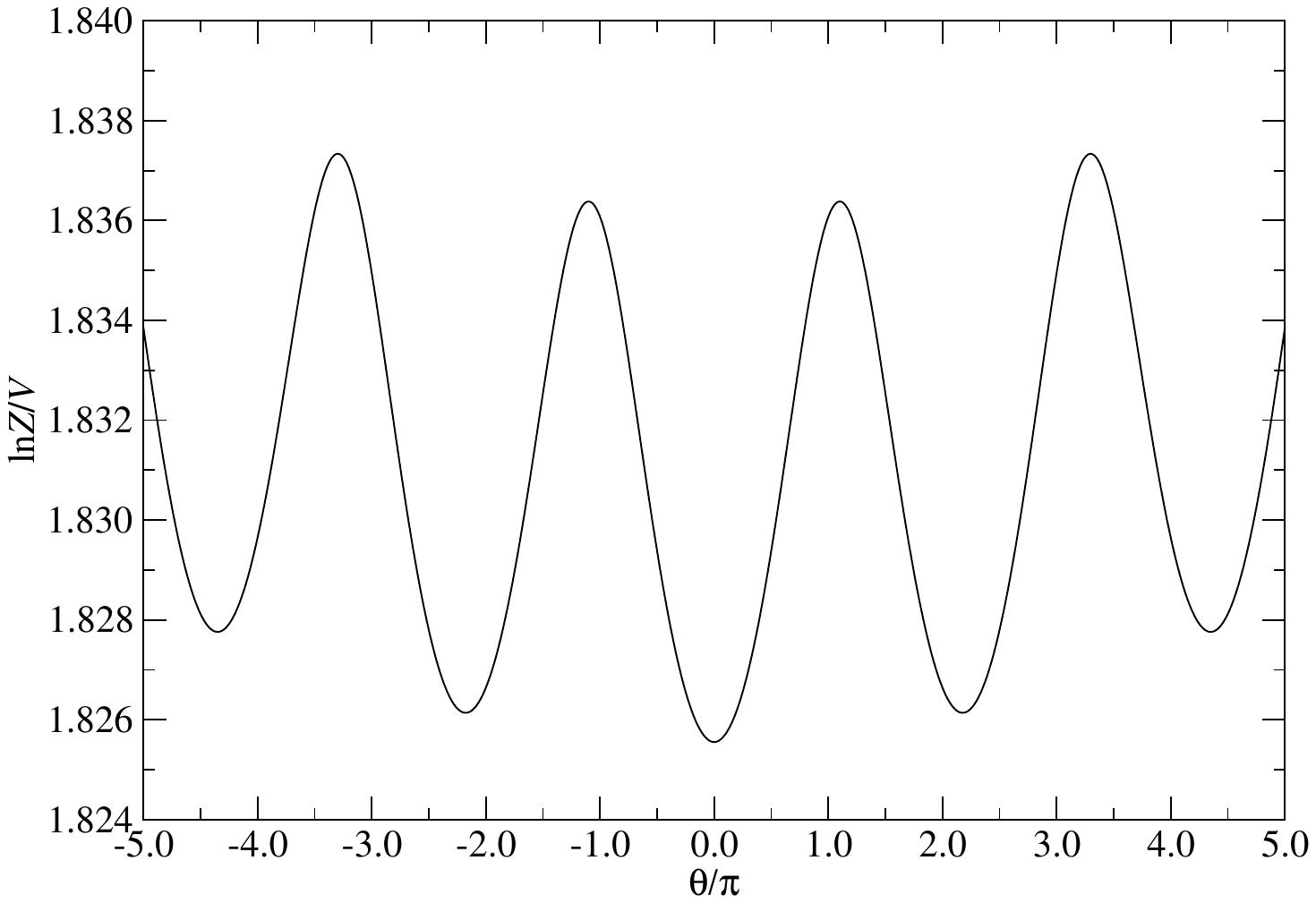}
        \end{minipage} 
    \end{tabular}
    \centering
    \text{\small (d) $(\beta,V)=(12.8,2^{7})$}\\
    \begin{tabular}{cc}
        \begin{minipage}[t]{0.43\hsize}
            \centering
            \includegraphics[width=1.0\hsize] {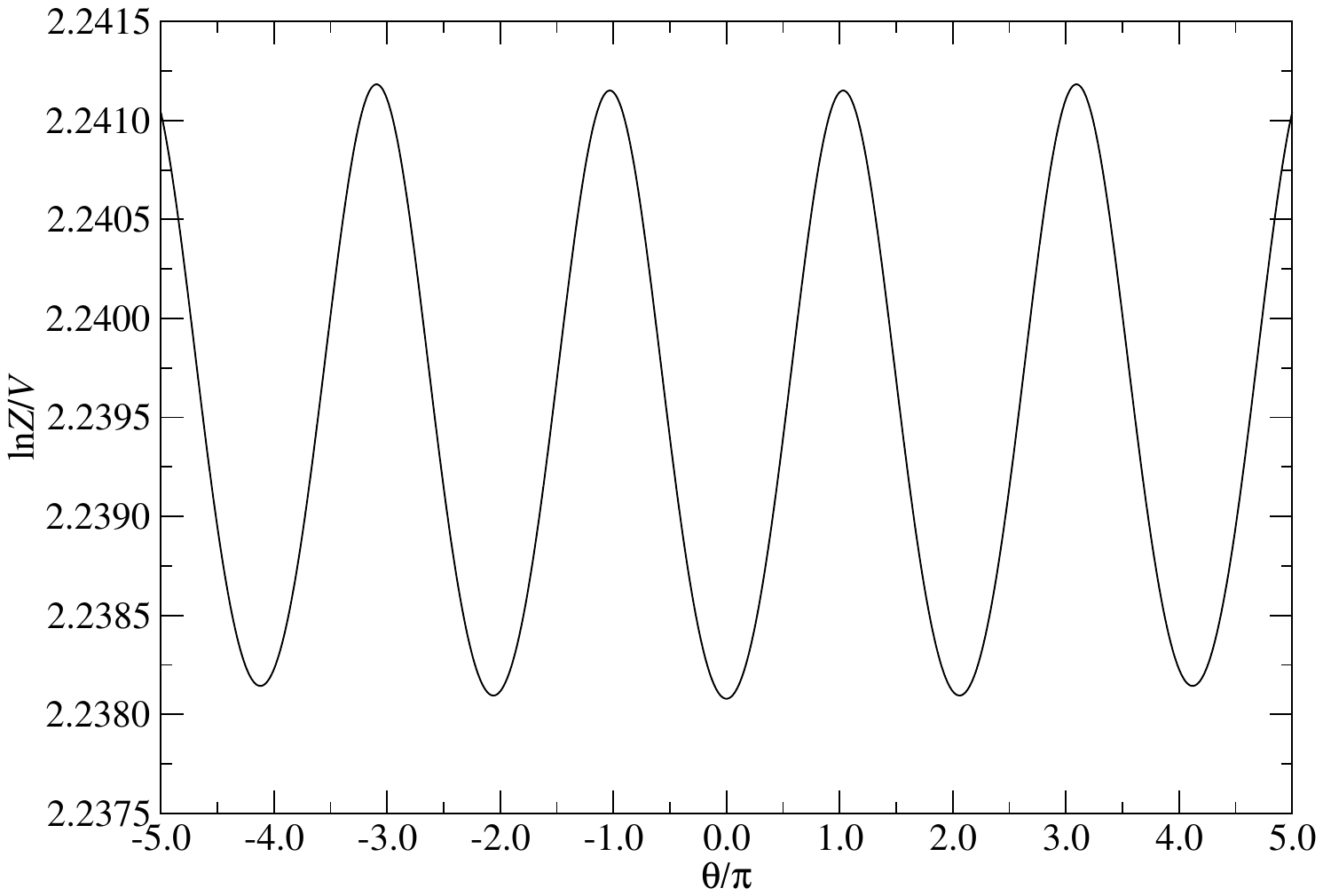} 
        \end{minipage} &
        \begin{minipage}[t]{0.43\hsize}
            \centering
            \includegraphics[width=1.0\hsize]{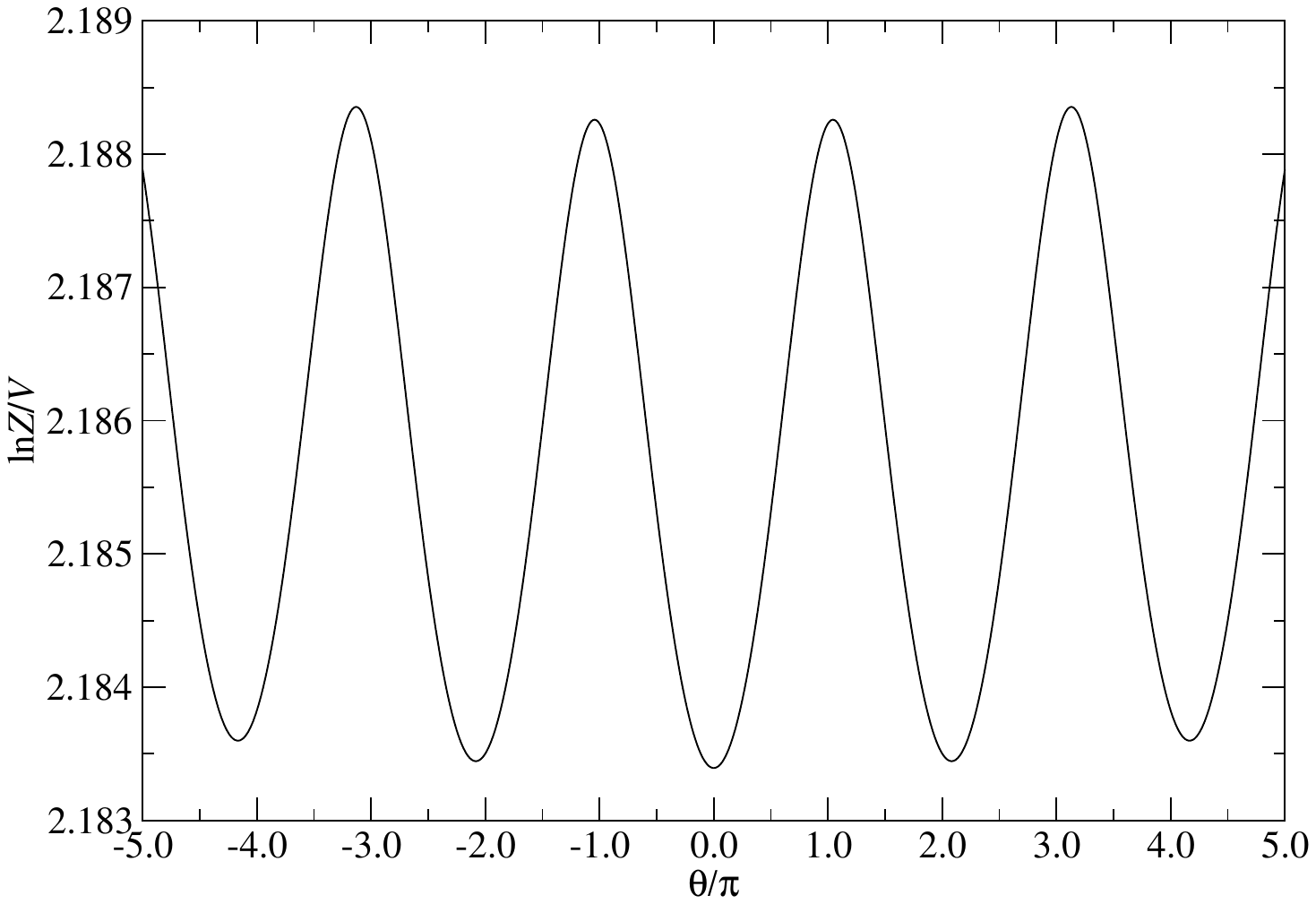}
        \end{minipage} 
    \end{tabular}
    \caption{Free energy density $\ln Z/V$ at various $(\beta,V)$ by the L{\"u}scher gauge action (left) and Wilson gauge action (right) with fixing $\beta/V=0.1$.}
    \label{fig:lnz_sin}
\end{figure}

\begin{figure}[htbp]
    \centering
    \text{\small (a) $(\beta,V)=(1.6,2^{4})$}\\
    \begin{tabular}{cc}
        \begin{minipage}[t]{0.43\hsize}
            \centering
            \includegraphics[width=1.0\hsize] {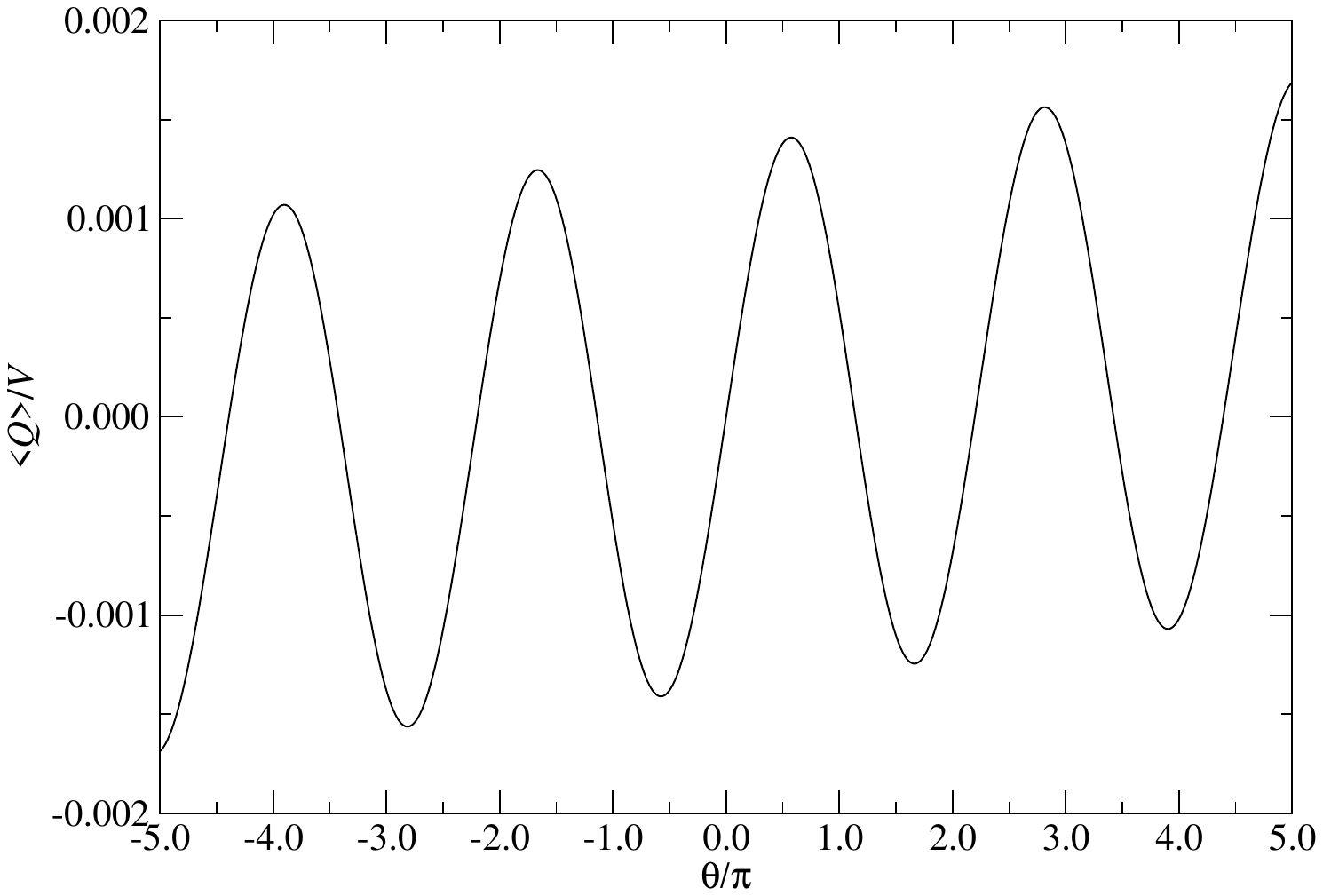} 
        \end{minipage} &
        \begin{minipage}[t]{0.43\hsize}
            \centering
            \includegraphics[width=1.0\hsize]{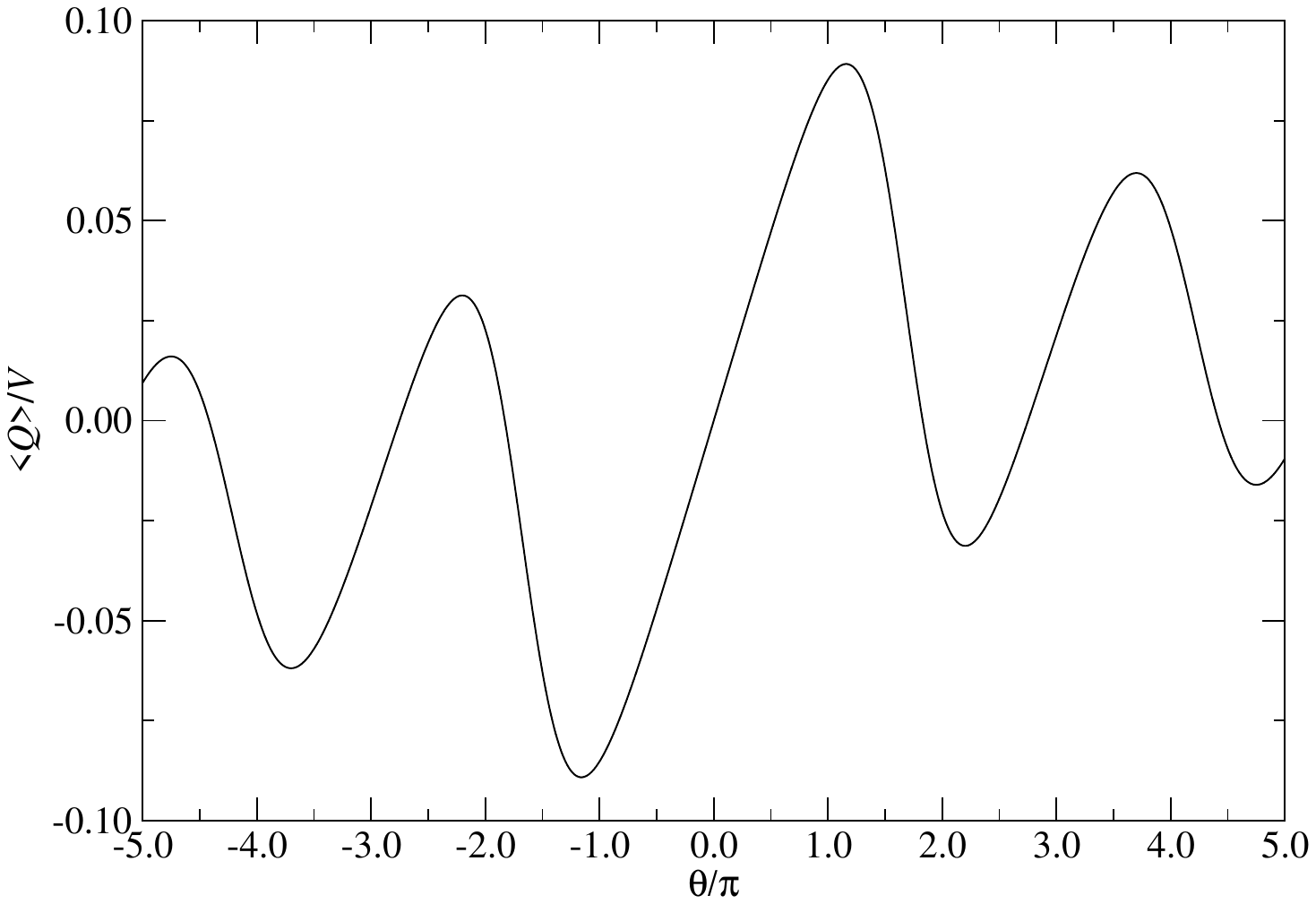}
        \end{minipage} 
    \end{tabular}
    \centering
    \text{\small (b) $(\beta,V)=(3.2,2^{5})$}\\
    \begin{tabular}{cc}
        \begin{minipage}[t]{0.43\hsize}
            \centering
            \includegraphics[width=1.0\hsize] {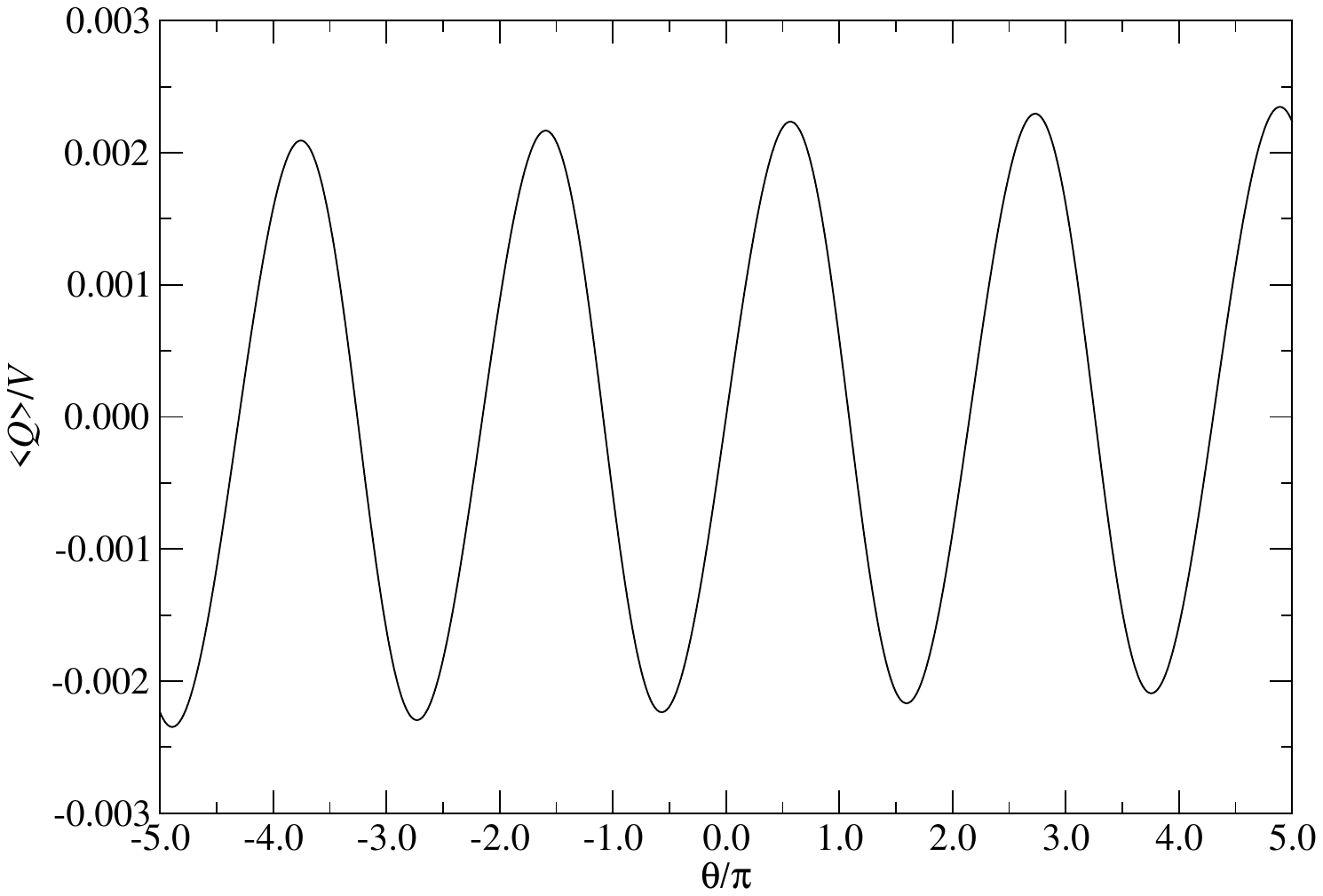} 
        \end{minipage} &
        \begin{minipage}[t]{0.43\hsize}
            \centering
            \includegraphics[width=1.0\hsize]{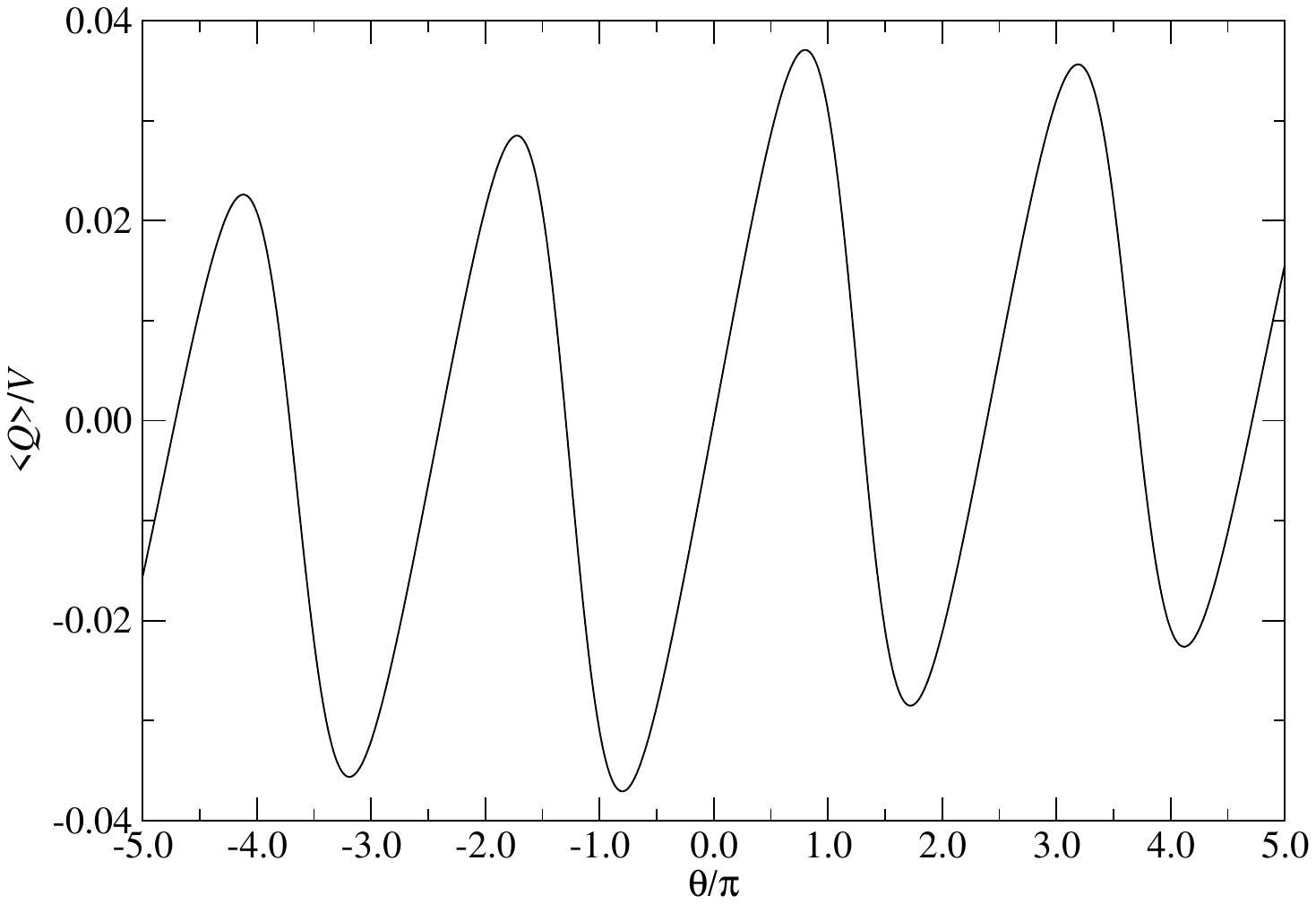}
        \end{minipage} 
    \end{tabular}
    \centering
    \text{\small (c) $(\beta,V)=(6.4,2^{6})$}\\
    \begin{tabular}{cc}
        \begin{minipage}[t]{0.43\hsize}
            \centering
            \includegraphics[width=1.0\hsize] {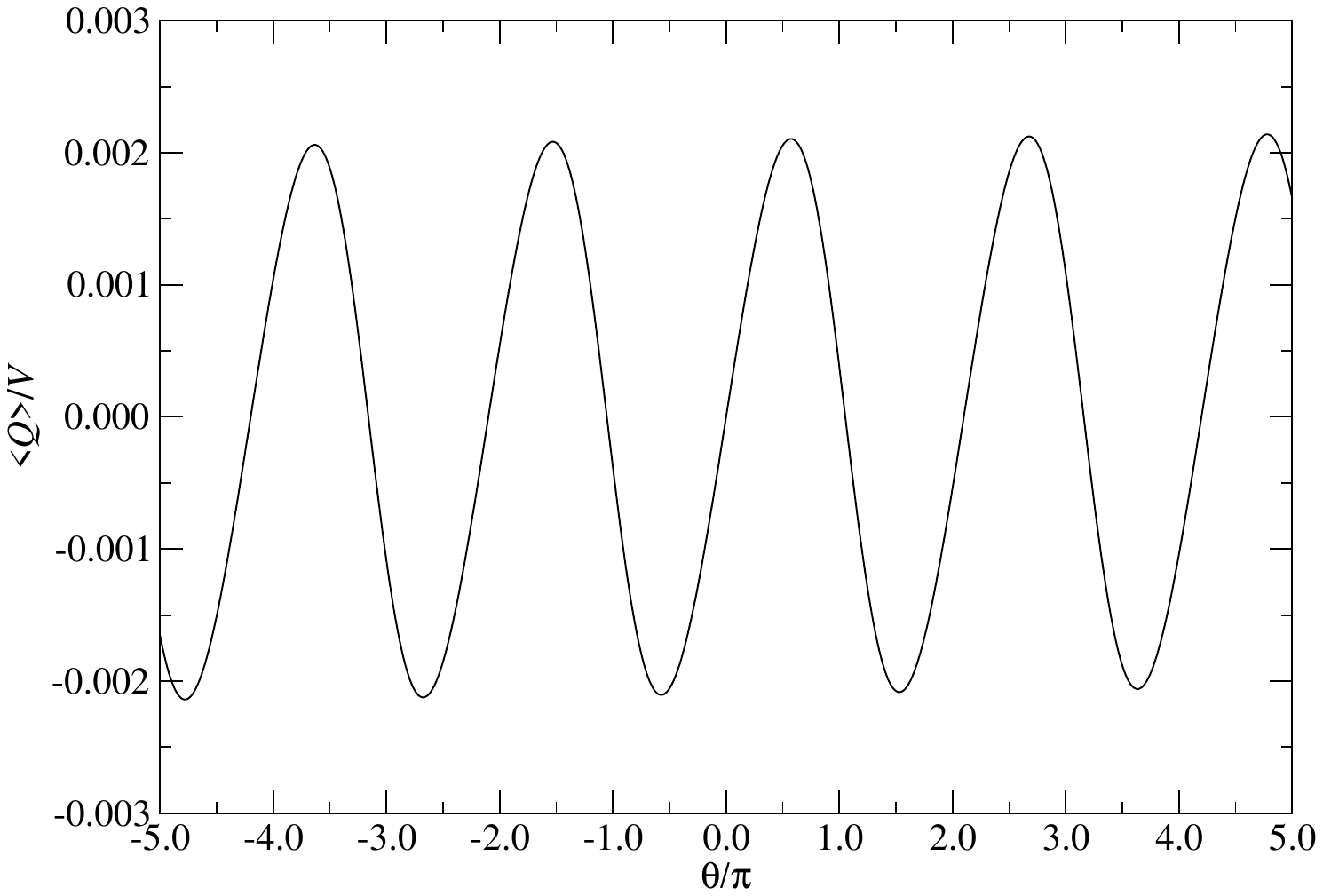} 
        \end{minipage} &
        \begin{minipage}[t]{0.43\hsize}
            \centering
            \includegraphics[width=1.0\hsize]{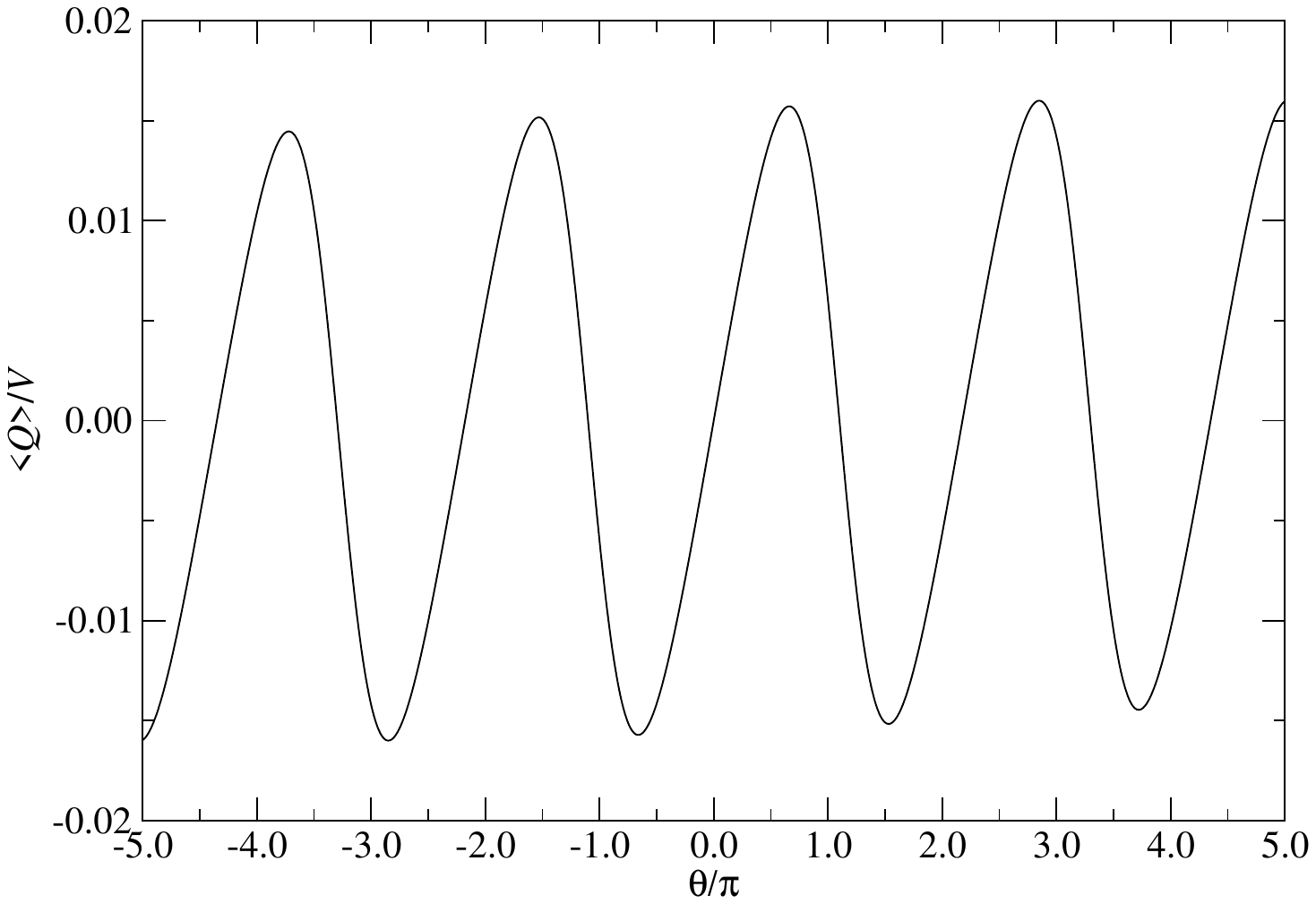}
        \end{minipage} 
    \end{tabular}
    \centering
    \text{\small (d) $(\beta,V)=(12.8,2^{7})$}\\
    \begin{tabular}{cc}
        \begin{minipage}[t]{0.43\hsize}
            \centering
            \includegraphics[width=1.0\hsize] {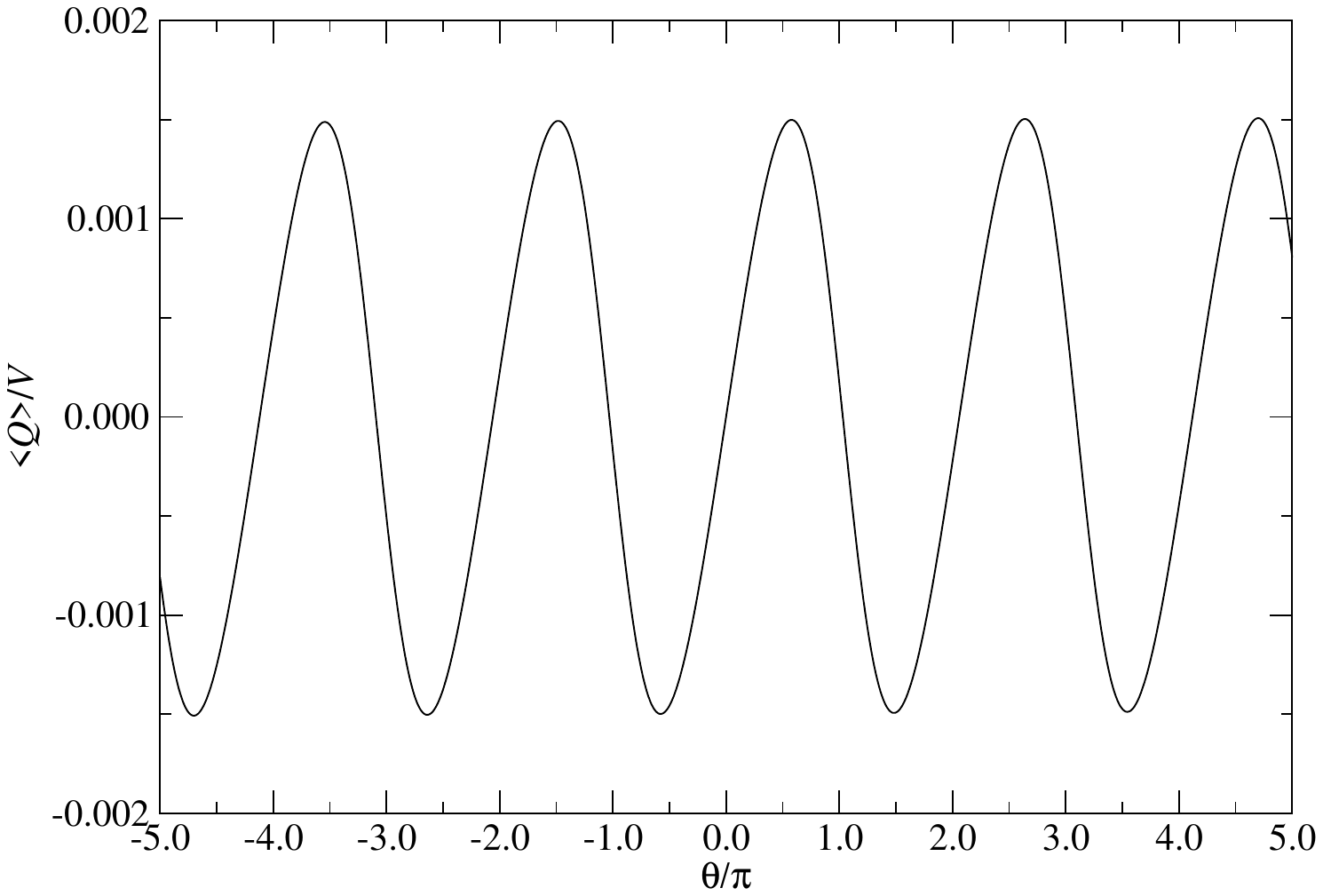} 
        \end{minipage} &
        \begin{minipage}[t]{0.43\hsize}
            \centering
            \includegraphics[width=1.0\hsize]{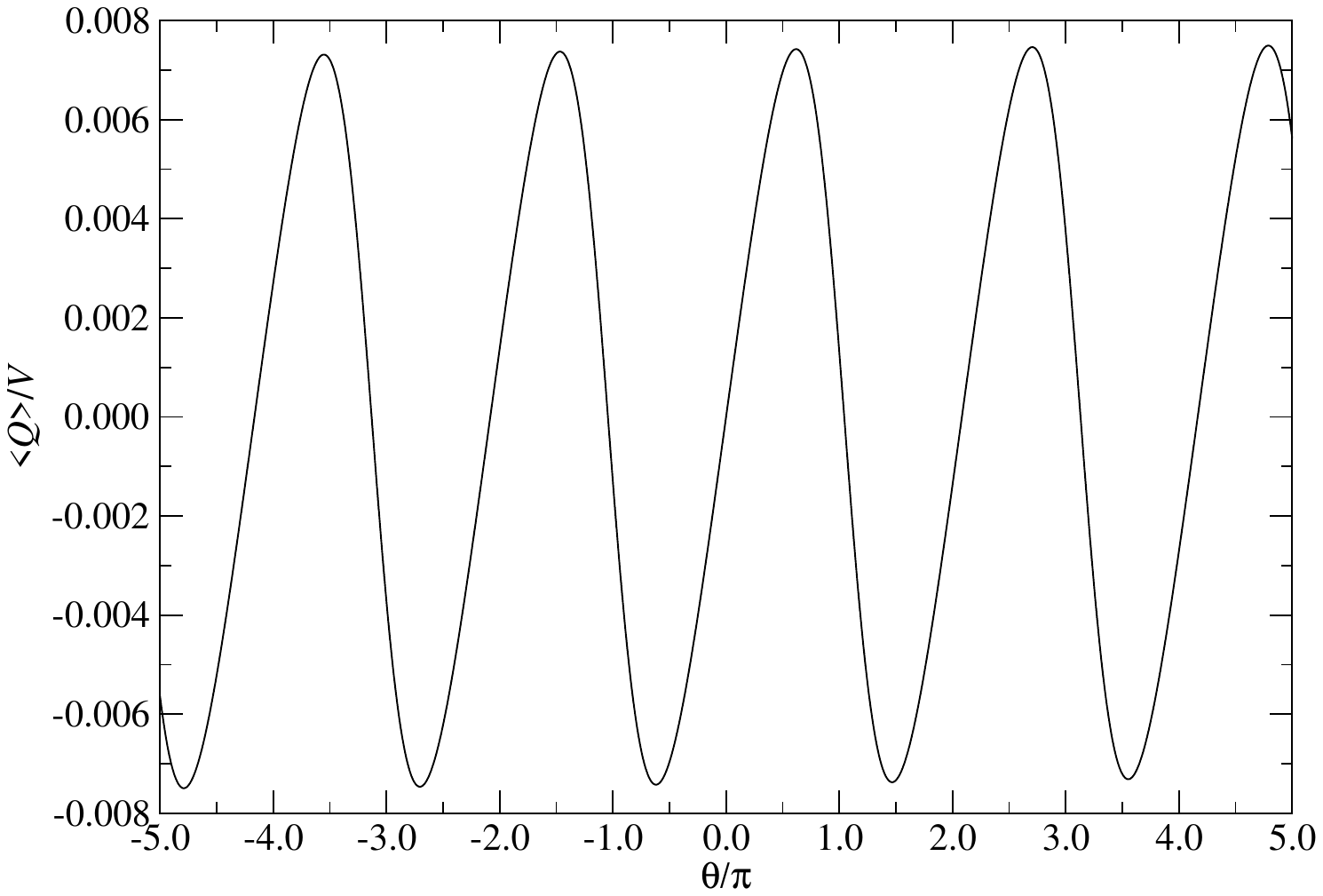}
        \end{minipage} 
    \end{tabular}
    \caption{Topological charge density $\langle Q\rangle/V$ at various $(\beta,V)$ by the L{\"u}scher gauge action (left) and Wilson gauge action (right) with fixing $\beta/V=0.1$.}
    \label{fig:top_charge_sin}
\end{figure}

Since the cutoff effects in the L{\"u}scher and Wilson gauge actions should differ, how to approach the continuum limit should also be different.
In Figure~\ref{fig:lnz_sin}, the free energy density $\ln Z/V$ is shown as a function of $\theta/\pi$ at various $\beta$ with fixing $\beta/V=0.1$.
All these results are obtained by the BTRG with $D=K_{g}=30$.
The left and right panels are obtained by the L{\"u}scher action with $\epsilon=1$ and Wilson action at the same $\beta$ and $V$.
Figure~\ref{fig:top_charge_sin} shows the topological charge density $\langle Q\rangle/V$ defined by Eq.~\eqref{eq:def_top} in the same way.
With the same $\beta$, the L{\"u}scher action gives a result closer to the continuum limit than the Wilson action.

For more quantitative discussion, we compute the topological susceptibility $\chi_{Q}$ given by Eq.~\eqref{eq:chi_top} employing the numerical differentiation with the $O(\Delta^{4})$ accuracy, setting $\Delta=\pi\times10^{-5}$.
In particular, we investigate the peak position of $\chi_{Q}$ varying $\beta$ and $V$ with fixing $\beta/V$, fitting $\chi_{Q}$ via three parameters $\chi_{\rm peak}$, $c$, and $\theta_{\rm c}$ by
\begin{align}
    \chi_{Q}(\theta)
    =
    \chi_{\rm peak}
    +
    c(\theta-\theta_{\rm c})^{2}
\end{align}
in the vicinity of the first peak appearing in the region where $\theta$ is positive.
Typically, we use 2000 data points around $\theta_{\rm c}$.
Table~\ref{tab:comparison_theta_c} summarizes the resulting peak position $\theta_{\rm c}$ in the unit of $\pi$.
At the same $\beta$, the L{\"u}scher action results in the peak position closer to $\theta_{\rm c}=\pi$ than the Wilson action.
Since we should observe the peaks in the topological susceptibility at $\theta=n\pi~(n=\pm1,\pm3,\cdots)$ in the limits of $\beta\to\infty$ and $V\to\infty$, we can conclude that the L{\"u}scher gauge action shows a faster convergence toward the continuum limit than the Wilson gauge action.

\begin{table}[htbp]
    \caption{
        Comparison of $\theta_{\rm c}/\pi$ obtained by the L{\"u}scher gauge action ($\epsilon=1$) and Wilson gauge action fixing $\beta/V=0.1$.
    }
    \begin{center}
        \begin{tabular}{ccc} \hline
            $\beta$ & L{\"u}scher ($\epsilon=1$) & Wilson \\ \hline
            $1.6$   & $1.11932(3)$               & 1.67903(2) \\
            $3.2$   & $1.08112(2)$               & 1.26026(1) \\
            $6.4$   & $1.05111(4)$               & 1.09604(7) \\
            $12.8$  & $1.03070(3)$               & 1.04296(1) \\ \hline
        \end{tabular}
    \end{center}
    \label{tab:comparison_theta_c}
\end{table}

\clearpage

\bibliographystyle{JHEP}
\bibliography{bib/formulation,bib/algorithm,bib/discrete,bib/grassmann,bib/continuous,bib/gauge,bib/real_time,bib/review,bib/for_this_paper}

\end{document}